\documentclass[11pt, a4paper]{article}
\pdfoutput=1

\usepackage{jheppub}
\usepackage[all]{hypcap}
\usepackage{multirow}
\usepackage{dsfont}

\newcommand{\br}{\textrm{BR}}
\newcommand{\sm}{\textrm{SM}}

\newcommand{\bsm}{\textrm{BSM}}
\newcommand{\unt}{\textrm{unt}}
\newcommand{\inv}{\textrm{inv}}
\newcommand{\tev}{\,\textrm{TeV}}
\newcommand{\gev}{\,\textrm{GeV}}
\newcommand{\mev}{\,\textrm{MeV}}

\newcommand{\m}{\,\textrm{m}}
\newcommand{\ns}{\,\textrm{ns}}
\newcommand{\mi}{\text{min}}
\newcommand{\ma}{\text{max}}
\newcommand{\Mt}{\ensuremath{\tilde{M}}}
\newcommand{\st}{\sin^2\theta}
\newcommand{\sint}{\ensuremath{\textrm{s}_{\theta}}}
\newcommand{\cost}{\ensuremath{\textrm{c}_{\theta}}}
\newcommand{\sinnot}{\ensuremath{\sin\left(\frac{\phi_0}{f}\right)}}
\newcommand{\sinnotsq}{\ensuremath{\sin^2\left(\frac{\phi_0}{f}\right)}}
\newcommand{\cosnot}{\ensuremath{\cos\left(\frac{\phi_0}{f}\right)}}

\newcommand{\mphi}{\ensuremath{m_{\phi}}}

\newcommand{\lint}{\mathcal{L}_{\textrm{int}} }

\newcommand{\fb}{\,\textrm{fb} }
\newcommand{\ab}{\,\textrm{ab} }
\newcommand{\iab}{\,\ab^{-1}}
\newcommand{\ifb}{\,\fb^{-1}}
\newcommand{\pb}{\,\textrm{pb} }

\newcommand{\cp}{\ensuremath{\mathcal{CP}}}

\newcommand{\sth}{s_{\theta}}

\newcommand{\eg}{\emph{e.g.}}
\newcommand{\ie}{\emph{i.e.}}

\newcommand{\lhp}{\ensuremath{\lambda_{h\phi}}}

\usepackage[table]{xcolor}
\newcommand{\leri}[1]{\left(#1 \right)}
\usepackage[caption=false]{subfig}

\usepackage{setspace}
\usepackage[section]{placeins}

\usepackage[nolist]{acronym}
\begin{acronym}
	\acro{SM}{Standard Model}
	\acro{BSM}{Beyond the Standard Model}
	\acro{NP}{New Physics}
	\acro{com}{center of mass}
	\acro{EFT}{effective field theory}
	\acro{DM}{Dark Matter}
	\acro{LHC}{Large Hadron Collider}
	\acro{TOF}{time of flight}
	\acro{ID}{inner detector}
	\acro{MS}{muon system}
	\acro{VBF}{vector boson fusion}
	\acro{EW}{electroweak}
	\acro{LO}{leading order}
	\acro{CL}{confidence level}
	\acro{NLO}{next to leading order}
	\acro{HL}{high luminosity}
	\acro{DD}{dynamic decoupling}
	\acro{EP}{equivalence principle}
	\acro{EWSB}{electroweak symmetry breaking}
	\acro{ALP}{Axion-Like Particle}
	\acro{HCAL}{hadronic calorimeter}
	\acro{ISR}{initial state radiation}
	\acro{DV}{displaced vertex}
	\acro{VEV}{vacuum expectation value}
\end{acronym}

\title{
Collider searches of scalar singlets across lifetimes
}
\author[a, b]{Elina Fuchs,}
\author[c]{Oleksii Matsedonskyi,}
\author[c]{Inbar Savoray,}
\author[a]{Matthias Schlaffer}

\affiliation[a]{University of Chicago, Department of Physics, 5720 South Ellis Avenue, Chicago, IL 60637 USA}
\affiliation[b]{Fermilab, Theory Department, Kirk Road and Pine Street, Batavia, IL 60510, USA}
\affiliation[c]{Weizmann Institute of Science, 234 Herzl Street, Rehovot 7610001, Israel}

\emailAdd{elinafuchs@uchicago.edu}
\emailAdd{oleksii.matsedonskyi@weizmann.ac.il}
\emailAdd{Inbar.Savoray@weizmann.ac.il}
\emailAdd{Schlaffer@uchicago.edu}

\preprint{FERMILAB-PUB-20-454-T,~EFI-20-17}

\abstract{ Spin-0 singlets arise in well-motivated extensions of the Standard Model.  Their lifetime
  determines the best search strategies at hadron and lepton colliders.  To cover a large range of
  singlet decay lengths, we investigate bounds from Higgs decays into a pair of singlets,
  considering signatures of invisible decays, displaced and delayed jets, and coupling fits of
  untagged decays.  We examine the generic scalar singlet and the relaxion, and derive a matching as
  well as qualitative differences between them. For each model, we discuss its natural parameter
  space and the searches probing it.  }

\begin{document}

\maketitle

\section{Introduction}
\label{sec:introduction}

Light spin-zero singlets are ubiquitous in models of \ac{NP}. They can have important
phenomenological roles such as serving as a portal to a Dark Sector~\cite{Patt:2006fw} and rendering
the electroweak phase transition first order to enable electroweak
baryogenesis~\cite{Anderson:1991zb,Espinosa:1993bs}.  In many cases, the phenomenology associated
with such NP can be encompassed in the minimal renormalizable extension of the \ac{SM} obtained by
adding one spin-zero singlet $\phi$~\cite{OConnell:2006rsp}.  We consider this model as a benchmark,
assuming all other new degrees of freedom are sufficiently heavy or weakly coupled to the \ac{SM}
particles.

Despite its simple setup, the singlet extension brings about a rich phenomenology related to the
Higgs, by opening the exotic decay channel $h\to\phi\phi$, if kinematically allowed (see
\eg~Ref.~\cite{Curtin:2013fra}), and by reducing the couplings of the Higgs boson to \ac{SM}
particles via singlet-Higgs mixing. This applies equally to scalars and pseudoscalars, though in
the latter case the $\phi$-Higgs mixing requires breaking of \cp{}.
The phenomenological implications reach far beyond Higgs-related observables, as the singlet
inherits the couplings of the Higgs to the \ac{SM} particles, suppressed by the mixing
angle. Therefore, the interactions of the singlet can, depending on its mass $m_\phi$, be probed
across the precision, luminosity and energy frontiers. The various signatures of the singlet include its effects on atomic physics, tests of
the equivalence principle, \ac{DM}, beam dump experiments, rare meson decays, collider
signatures as well as astrophysical and cosmological observables, see
\eg~Refs.~\cite{Berengut:2017zuo,Aharony:2019iad,Antypas:2019qji,McDonald:1993ex,Burgess:2000yq,Urena-Lopez:2019kud,Beniwal:2017eik,Hardy:2016kme,Clarke:2013aya,Dolan:2014ska,Flacke:2016szy,Choi:2016luu,Barger:2007im,Robens:2016xkb,Heinemann:2019trx,Huang:2016cjm,Beacham:2019nyx}
and references therein. In this work, we focus on the collider searches in the mass range of
$3\gev\leq m_\phi \leq m_h/2$.

In addition to the above generic renormalizable extension, we consider the relaxion
framework~\cite{Graham:2015cka}, which offers an alternative approach to the gauge hierarchy problem,
and can also provide a \ac{DM} candidate~\cite{Fonseca:2018kqf,Banerjee:2018xmn}, facilitate
baryogenesis~\cite{Gupta:2019ueh}, and address other hierarchies of the SM~\cite{Davidi:2018sii}.
The relaxion is a naturally light pseudo-Nambu-Goldstone field, whose variation in the early Universe
induces the scanning of the Higgs mass.  The key ingredient of the relaxion mechanism is the
so-called backreaction potential, which stops the relaxion evolution at the observed Higgs mass
value. The backreaction potential is responsible for the interactions between the Higgs and the
relaxion which are relevant for the collider phenomenology, realizing the Higgs portal structure
similarly to the generic singlet extension discussed above~\cite{Flacke:2016szy,Choi:2016luu}. The
mass range examined in this work, however, represents only the extremely heavy region of the full
relaxion parameter space.

The two main parameters in our focus are the Higgs-relaxion mixing and the $h^2\phi^2$ coupling.  We
demonstrate that, while having many similarities with the generic portal models, the relaxion is
more constrained, but at the same time allows for larger values of the mixing angle than in the
generic portal scenarios. This feature occurs because the naturalness considerations, which can be
used to constrain the portal parameter space, can be automatically violated by the dynamics of relaxion
field~\cite{Banerjee:2020kww}.

While collider constraints on promptly decaying relaxions or light scalar singlets were derived in
Ref.~\cite{Frugiuele:2018coc}, here we focus on the range of couplings that makes the scalar
sufficiently long-lived such that it does not decay promptly. We take indirect constraints from
global Higgs coupling fits as well as direct searches for invisible Higgs decays and displaced and
delayed signatures into account. Moreover, we investigate the implications of the bounds on untagged
Higgs decays on singlets decaying in the detector. For each method, we evaluate the potential of
various hadron and lepton colliders to probe natural parameter space.

The paper is structured as follows. In Sec.~\ref{sec:higgs-port-phen} we present general properties
of the singlet extension. In particular, we consider the renormalizable singlet in
Sec.~\ref{sec:general-singlet-extension} and the special case of the relaxion in
Sec.~\ref{sec:relaxion-as-special}. Sec.~\ref{sec:collider-bounds-long} contains the collider
bounds ordered by the scalar lifetime. We compare these complementary bounds in
Sec.~\ref{sec:overview} before concluding in Sec.~\ref{sec:conclusion}.

\section{Singlet extension of the Higgs sector}
\label{sec:higgs-port-phen}

In the following we present the phenomenological features of the real scalar singlet extension of the
\ac{SM} Higgs sector. After discussing the general properties of the scalar, we derive the masses and the relevant couplings for the renormalizable
$\mathds{Z}_2$ breaking model, as well as for the non-renormalizable relaxion framework. This direct
comparison will allow us to make a mapping from one model to the other.

The most general extended scalar potential of the Higgs doublet $H$ and a new singlet scalar $\Phi$
is given by
\begin{equation}
  \label{eq:general-lagrangian}
  V_\mathrm{s}(\Phi, H) = V(\Phi) + \mu^2(\Phi) H^\dagger H + \lambda_h
  \left(H^\dagger H\right)^2\,,
\end{equation}
where $V(\Phi)$ and $\mu^2(\Phi)$ are functions of the field $\Phi$, whose exact forms depend on the
model.  We do not consider direct couplings between $\Phi$ and other \ac{SM} states besides the
Higgs. In general, both fields can be written in the unitary gauge as
\begin{align}
  \label{eq:vevs}
  H = \left(0, \frac{v+h}{\sqrt{2}}\right)^T\,, \quad \Phi = \phi_0+\phi\,,
\end{align}
where $v=246\gev$ and $\phi_0$ are their respective \acp{VEV}, and their
dynamical degrees of freedom are denoted by $h$ and $\phi$.

If $\Phi$ is not protected by an unbroken $\mathds{Z}_2$ symmetry, the singlet $\phi$ mixes with the
Higgs. In this case, $\phi$ inherits the Higgs couplings to the \ac{SM} particles, suppressed by the
mixing angle $\sin\theta$. At the same time, the couplings of the Higgs boson to the \ac{SM} particles
are reduced by a global factor $\cos \theta$. The mass matrix elements are
\begin{align}
  \label{eq:massmatrix}
  m^2_{hh} = 2 v^2 \lambda_h, \quad m^2_{\phi\phi} = V''(\phi_0) + \frac{1}{2} v^2
  \mu^{2\prime\prime}(\phi_0),\quad m^2_{h\phi} = v  \mu^{2\prime}(\phi_0)\,,
\end{align}
where the derivative is with respect to $\phi$. Defining the mixing angle by
\begin{equation}
  \label{eq:rotation}
  \begin{pmatrix}
    \phi_\textnormal{phys.}\\h_\textnormal{phys.}
  \end{pmatrix}
  =
  \begin{pmatrix}
    \cos\theta & - \sin\theta\\
    \sin\theta & \cos\theta
  \end{pmatrix}
  \begin{pmatrix}
    \phi\\h
  \end{pmatrix},
\end{equation}
for $m_{hh}^2 \gg m_{\phi\phi}^2,\ m_{h\phi}^2$, it can be approximated by
\begin{equation}
\sin \theta \approx \frac{m_{h\phi}^2}{m_{hh}^2} =  \frac{\mu^{2\prime}(\phi_0)}{2\lambda_h v}\,.
\end{equation}
Since for small mixing angles, the mass and interaction eigenstates are approximately the same, we
use the symbols $h$ and $\phi$ for both states throughout the paper and drop the label of the
physical mass eigenstates used above.  We denote the physical masses by $m_\phi$ and $m_h$,
respectively, and their expressions will be given in the following sections. Due to the mixing, the
$\phi$ production from and decay into \ac{SM} particles equal those of a \ac{SM} Higgs boson with
the respective $\phi$ mass, modified by the mixing angle.

In addition, if the mixing angle or $\mu^{2\prime\prime}(\phi_0)$ are
non-zero\footnote{For $\phi_0=0$ this requires that $\mu^2$ contains a $\phi^2$ term whereas for
  $\phi_0\neq 0$ higher powers of $\phi$ are also a valid solution.  These terms can be explicit or
  can arise from an expansion in $\phi$.}, the Higgs couples to a pair of singlets. We denote this
coupling of the mass eigenstates by $c_{h\phi\phi}$, which receives the following contributions
\begin{align}
  \label{eq:c_hphiphi}
  c_{h\phi\phi} &= 3 \cost \sint^2 v \lambda_h
                  + \left(\frac{1}{2} \sint^3 - \cost^2 \sint\right)\mu^{2\prime}(\phi_0)
                  + \left(\frac{1}{2} \cost^3 - \cost \sint^2\right) v \mu^{2\prime\prime}(\phi_0)\nonumber \\
                &\quad  + \frac{1}{4} \cost^2 \sint v^2 \mu^{2\prime\prime\prime}(\phi_0)
                  + \frac{1}{2} \cost^2 \sint V^{\prime\prime\prime}(\phi_0)
                  \,,
\end{align}
where we use the shorthand notation $\sint\equiv\sin\theta$ and $\cost\equiv\cos\theta$. In the two concrete
models considered below, this expression simplifies substantially, especially in the limit of small
mixing. When $ 2 m_\phi< m_h$, the Higgs can decay via the new channel $h\to \phi\phi$ with a decay
width of
\begin{align}
  \label{eq:decay_width}
  \Gamma_{h\to\phi\phi} = \frac{c_{h\phi\phi}^2}{8 \pi m_h}
  \sqrt{1-\frac{4m^2_\phi}{m^2_h}}\,.
\end{align}

\subsection{Renormalizable singlet}
\label{sec:general-singlet-extension}

The most general renormalizable form of $V(\Phi)$ and $\mu^2(\Phi)$ is
\begin{align}
  \label{eq:singlet-extension}
  V(\Phi) & = t \Phi + \frac12 m_0^2 \Phi^2  + \frac{a_\phi}{3}\Phi^3 + \frac{\lambda_\phi}{4} \Phi^4\,,\\
  \mu^2(\Phi) & = -\mu_0^2  + 2 a_{h\phi} \Phi +\hat{\lambda}_{h\phi} \Phi^2\,.
\end{align}
This theory, often also called a \emph{Higgs portal model}, has been studied extensively in the
literature, see
\eg{}~\cite{Piazza:2010ye,Robens:2016xkb,Ilnicka:2018def,Heinemann:2019trx,Huang:2016cjm,Dawson:2017vgm,Kotwal:2016tex}
and references therein. For later convenience, we choose
$\phi_0=0$, which can always be obtained by a shift of the $\phi$ field. This implies $t = -a_{h\phi}
v^2$ from the minimum condition of the scalar potential $V_\text{s}$. In this model the mixing
angle is
\begin{align}
  \label{eq:mix_singlet}
  \sin \theta = \frac{1}{\sqrt{2}}\sqrt{1-\frac{1}{\sqrt{1+x_\text{int}^2}}} = \frac{1}{\sqrt{2}}
  \sqrt{1-\sqrt{1-x_\text{phys}^2}}\approx \frac{a_{h\phi}}{v \lambda_h}\,,
\end{align}
where $x_\text{state}=4 v a_{h\phi}/\Delta m_\text{state}^2$, with
$\Delta m^2_\text{int}= m_{hh}^2-m_{\phi\phi}^2$ being the difference of the diagonal entries of the
mass matrix before diagonalization, and $\Delta m^2_\text{phys}= m_{h}^2-m_{\phi}^2$ being the
difference of the physical mass eigenvalues\footnote{Hence the difference in the squared physical
  masses can be expressed as
  $\Delta m_\text{phys}^2 = \Delta m_\text{int}^2 \sqrt{1+x_\text{int}^2}$.}. The approximation in
the last step holds for $\hat{\lambda}_{h\phi} v^2 + m_0^2 \ll 2 \lambda_h v^2$ and
$ a_{h\phi}\ll v \lambda_h$. This corresponds to a large mass splitting between the singlet and the
Higgs and a small mixing angle. The physical masses are
\begin{align}
  \label{eq:mass_singlet}
  m^2_{\phi, h} = \frac{1}{2}\left(m_0^2 + v^2(2\lambda_h + \hat{\lambda}_{h\phi}) \mp v^2
  \sqrt{\left(\frac{4 a_{h\phi}}{v}\right)^2+ \left(\frac{m_0^2}{v^2} + \hat{\lambda}_{h\phi}-2\lambda_h\right)^2}\right)\,.
\end{align}
For $\left|a_{h\phi}\right|\ll v \lambda_h$ the masses are approximated as
\begin{align}
  \label{eq:mass_singlet_approx_mphi}
  m_\phi^2 &\approx m_0^2 + v^2\hat{\lambda}_{h\phi} - \Delta_m\\
  \label{eq:mass_singlet_approx_mh}
  m_h^2 &\approx 2 v^2 \lambda_h + \Delta_m\,\qquad \mathrm{with}\\
  \label{eq:mass_singlet_approx_deltaM}
  \Delta_m &= \frac{4a_{h\phi}^2}{2\lambda_h - m_0^2/v^2 - \hat{\lambda}_{h\phi}}
             \approx 2 v^2 \sint^2 \lambda_h
             \approx m_h^2 \sint^2\,,
\end{align}
where the approximations rely on a small mixing and a large splitting of the diagonal entries of the
mass matrix, exactly as the approximation in Eq.~\eqref{eq:mix_singlet}. Using in
Eq.~\eqref{eq:c_hphiphi} the explicit expressions for $V(\Phi)$ and $\mu^2(\Phi)$ given in this
section, we obtain the explicit coupling $c_{h\phi\phi}$
\begin{align}
  \label{eq:chphiphi_singlet}
  c_{h\phi\phi} &= 3 \cost\sint^2 v\lambda_h
                  + a_{h\phi} (\sint^3 - 2\cost^2\sint)
                  + \hat{\lambda}_{h\phi}v (\cost^3-2\cost \sint^2)
                  + a_\phi \cost^2\sint\\
  \label{eq:chphiphi_singlet_approx}
                &\approx \sint^2 v\lambda_h
                  + \hat{\lambda}_{h\phi}v
                  + a_\phi \sint\\
  \label{eq:chphiphi_singlet_approx2}
                &\approx \sint^2 \frac{m_h^2}{2v}
                  + \lambda_{h\phi}v \,,
\end{align}
where the approximation holds for small mixing and makes use of Eq.~\eqref{eq:mix_singlet}, and we
define
\begin{equation}
  \label{eq:lambdahphi}
  \lambda_{h\phi} \equiv \hat{\lambda}_{h\phi} + a_\phi\frac{\sint}{v}\,.
\end{equation}
We use this as a parameter in the phenomenological investigations.

\subsubsection*{Theoretical bounds on the parameter space}
\label{sec:naturalness}

The relevant phenomenology is described by the four physical parameters $m_\phi$, $\sth$,
$\hat{\lambda}_{h \phi}$, and $a_\phi$. The parameters \sint{}~\cite{Piazza:2010ye,Graham:2015ifn},
$\hat{\lambda}_{h\phi}$, and $a_\phi$ contribute to $m_\phi$, the former two at tree-level and
$a_\phi$ via a $\phi$-loop. Therefore, their viable ranges are bounded by naturalness and depend on
$m_\phi$ as\footnote{We here neglect the $\log \Lambda$ dependence of the upper limit on $a_\phi$.}
\begin{align}
  \sin\theta &\lesssim \frac{m_\phi}{m_h} \,,\label{eq:genmixmax}\\
  \hat{\lambda}_{h \phi} &\lesssim \frac{m_\phi^2}{v^2}\,,\\
  a_\phi &\lesssim 4\pi m_\phi\,. \label{eq:aphimax}
\end{align}
The upper naturalness bound on $\lambda_{h\phi}$ is then given by
\begin{equation}\label{eq:lhphimax}
 \lambda_{h\phi}^{\rm{max}}=\frac{m_\phi^2}{v^2} + 4\pi \frac{m_{\phi}}{v}\sint\,.
\end{equation}

As we will see in the specific case of the relaxion, such naturalness bounds may be violated by
orders of magnitude as a consequence of the cosmological evolution of the fields.

\subsection{Relaxion}
\label{sec:relaxion-as-special}

Unlike the generic Higgs portal model considered above, the relaxion scenario is designed to solve
the \ac{SM} hierarchy problem and is therefore much more constrained and predictive.  First, we
briefly summarize the cosmological relaxation mechanism~\cite{Graham:2015cka}, considering the
relaxion potential of the form
\begin{align}
   V(\Phi) &=  r g \Lambda^3 \Phi\,,\label{eq:Vsr}\\
  \mu^2(\Phi) & = -\Lambda^2 + g \Lambda \Phi - \Mt^2 \cos\left(\frac{\Phi}{f}\right)\,. \label{eq:mu2relax}
\end{align}
Here $\Lambda$ is a UV cutoff, $\Mt$ is the height of the backreaction potential (see below) and $f$
is the relaxion oscillation scale. During its evolution, the relaxion scans the Higgs mass parameter
$\mu^2(\Phi)$ from a large and positive value $\sim \Lambda^2 \gg v^2$ down to negative values. This
scanning is a result of the slow-roll potential $V(\Phi)$, which is controlled by the small
dimensionless coupling $g$, and $r > {1} /{16 \pi^2}$ which is bounded from below by the
requirement of technical naturalness \cite{Gupta:2015uea}. Once $\mu^2(\Phi)$ becomes negative, the
Higgs gets a \ac{VEV} and thereby activates a backreaction potential $\propto \cos (\Phi/f)$, which
eventually stops the rolling of the relaxion at a value $\phi_0$, where $v(\phi_0)=246\gev$ (see \cite{Fonseca:2018xzp} for a recent discussion of the stopping mechanisms).

Such a theory naturally generates a large hierarchy between the electroweak scale and $\Lambda$,
solving the \ac{SM} naturalness problem\footnote{The relaxion does not solve the gauge hierarchy
  problem up to the Planck scale, and thus requires a UV completion to provide the needed
  $\Lambda\ll M_\textnormal{Pl}$~\cite{Batell:2015fma,Evans:2016htp,Batell:2017kho}, and also to
  produce a large relaxion field excursion~\cite{Choi:2015fiu,Kaplan:2015fuy}.}. In the following,
we require
\begin{equation}
f \geq \Lambda \geq \Lambda_{\text{min}} = 1\, \text{TeV} \,.
\end{equation}
The backreaction mechanism is model-dependent, and its most general potential is
\begin{eqnarray} 
  V_\text{br}( h,\phi)= -\tilde M^{4-j} \left(\frac{v+h}{\sqrt{2}}\right)^j \cos\left(\frac{\phi}{f}\right),
   \label{eq:br}
\end{eqnarray} 
where we chose $j=2$ and assume the minimal scenario of~\cite{Graham:2015cka} (for alternative
scenarios see
\eg{}~Refs.~\cite{Graham:2015cka,Nelson:2017cfv,Hook:2016mqo,Espinosa:2015eda,Gupta:2015uea,Davidi:2017gir}).
To suppress Higgs-independent loop-induced corrections to the backreaction
potential~\cite{Espinosa:2015eda}, the backreaction scale has to satisfy
$\tilde M^2 \ll 8 \pi^2 v^2$. Concretely, we require
 \begin{equation}
\tilde M \leq  \tilde M_{\text{max}} = 1\tev
 \label{lambdabrMAX}\,.
\end{equation}

\subsubsection{Comparison to singlet extension}
\label{sec:comp-singl-extens}

Around a local relaxion minimum $\langle\phi\rangle=\phi_0$, all the phenomenologically relevant
features of the relaxion model can be derived from those of the singlet extension discussed in
Sec.~\ref{sec:general-singlet-extension}, by substituting
\begin{align}
  \label{eq:relaxtolin}
  m^2_0,\, a_\phi,\, \lambda_\phi &  \to 0 \,,\\
  a_{h\phi} & \to \sinnot \frac{\Mt^2}{2f} + \frac{g\Lambda}{2}\,,\\
              \hat{\lambda}_{h\phi} & \,\to\, \cosnot \frac{\Mt^2}{2f^2} \,.
\end{align}
Making these substitutions in Eqs.~\eqref{eq:mix_singlet} and~\eqref{eq:mass_singlet_approx_mphi},
and omitting the term suppressed by the small coupling $g$, we obtain
\begin{align}
  \label{eq:relaxion_pheno_from_theory_parameters_sint}
  \sint & \approx  \frac{\Mt^2}{2vf\lambda_h} \sinnot \,,\\
  \label{eq:relaxion_pheno_from_theory_parameters_mphi}
  m_\phi^2 & \approx \frac{v^2 \tilde M^2}{2 f^2} \cosnot - \frac{v^2}{m_h^2} \frac{\Mt^4}{f^2}
          \sinnotsq\,,
\end{align}
where we neglect small corrections to the Higgs mass $m_h$. We notice that all the other couplings can
be expressed as functions of $m_\phi$ and $\sint$ as
\begin{align}
  \hat{\lambda}_{h \phi}  = \lambda_{h \phi} &= \frac{m_\phi^2}{v^2} + \frac{m_h^2}{v^2}\sint^2\,,\label{eq:lhpRel}\\
  a_{h\phi} &=\frac{m_h^2}{2v} \sint\,.
\end{align}
This means that this relaxion model has only two free parameters relevant for collider
phenomenology, \ie{}~two less than the generic singlet case. The triple scalar coupling
$c_{h\phi\phi}$ can then be written as
\begin{align}
  \label{eq:c_hphiphi_relaxion}
  c_{h\phi\phi} &\approx \frac{m_\phi^2}{v} + \frac{3}{2} \frac{m_h^2}{v}\sint^2\,.
\end{align}

Hence, in contrast to the renormalizable singlet extension that has $\hat{\lambda}_{h\phi}$ and
$a_{h\phi}$ as additional parameters, in the relaxion model this coupling is fully determined by
$m_\phi$ and $\sint$. Thus, the viable phenomenological parameter space is more limited and the model is more predictive.

\subsubsection{Theoretical bounds on the parameter space}
\label{sec:RelaxionBand}
Naively, the general naturalness bound on $\sint$ obtained in Eq.~\eqref{eq:genmixmax} applies also
to the relaxion model. However, following Refs.~\cite{Banerjee:2018xmn,Banerjee:2020kww}, the
dynamical evolution of the relaxion can fix the value of $\phi_0$ at such a position that the two
contributions to the relaxion mass in Eq.~\eqref{eq:relaxion_pheno_from_theory_parameters_mphi}
cancel each other to a high precision, leading to a larger allowed value for $\sin \theta$ for a
given mass. In the following, we denote the number of a minimum by $n$.

\paragraph{First minimum}
The degree of such a cancellation is maximal in the first local minimum of the relaxion
potential. There, in the limit of $\tilde M \gg \sqrt{\lambda_h}v$ , the relaxion mass and mixing
angle are given by (see Appendix~\ref{sec:relaxstop})\footnote{Obtaining $m_\phi$ in the GeV range
  necessitates a large value of $\Mt$, and therefore the limit $\tilde M \gg \sqrt{\lambda_h}v$ is
  justified, see Eq.~\eqref{eq:relaxion_pheno_from_theory_parameters_mphi}.}
\begin{align}
m_\phi^2  &\approx \sqrt{\frac{3 \pi}{2 \lambda_h^{1/2}}} \frac{(v \tilde M)^{5/2}}{f^2 \Lambda}\,, \label{eq:relaxmass1min}\\
\sin \theta &\approx \frac{\tilde M}{2 \sqrt \lambda_h f}\,. \label{eq:relaxmix1min}
\end{align}
The mixing is maximized for maximal $\tilde M$ and  minimal $f$, namely $f=\Lambda$.  Expressing
$f$ in terms of $m_\phi$ from Eq.~\eqref{eq:relaxmass1min} and substituting this in
Eq.~\eqref{eq:relaxmix1min} yields
\begin{equation}\label{eq:mixmax}
  \sin \theta <  \left(\frac{\Mt_{\textnormal{max}}}{96 \pi v \lambda_h^{5/2}}\right)^{1/6} \left(\frac{m_\phi}{v}\right)^{2/3}\,.
\end{equation}
Thus, the mixing is parametrically enhanced, as it is proportional to $ (m_\phi/v)^{2/3}$ instead of
the naturally expected $\propto m_\phi/v$, with the prefactor of $\mathcal{O}(1)$.  A relaxion with
a larger mixing than that defined in Eq.~(\ref{eq:mixmax}) corresponds to an \emph{unnatural} tuning
of the relaxion mass.

Solving Eq.~\eqref{eq:relaxmass1min} for $f$, substituting it in Eq.~(\ref{eq:relaxmix1min}), and
setting $\Lambda$ ($\Mt$) to its minimal (maximal) value, we obtain the lower bound on the mixing
angle
\begin{equation}\label{eq:mixmin}
\sin \theta \simeq\left(\frac{1}{24\pi\lambda_h^{3/2}}\right)^{1/4} \frac{m_\phi \Lambda^{1/2}}{v^{5/4} \tilde M^{1/4}} >
\left(\frac{\Lambda_{\text{min}}^2}{24\pi\lambda_h^{3/2}v \Mt_{\text{max}}}\right)^{1/4}
\frac{m_\phi}{v} \approx \frac{m_\phi}{v}\,.
\end{equation}

\paragraph{Generic minimum}
As mentioned above, the degree of tuning decreases if the relaxion stops in a later minimum. This
may happen either through quantum fluctuations or by classical
rolling~\cite{Fonseca:2019ypl,Fonseca:2019lmc}. In the limit of small tuning in a far minimum,
$n\gg1$, $\sin(\phi_0/f)\sim\cos(\phi_0/f)\sim\mathcal{O}(1)$ and naturalness arguments lead to an
estimate of the minimal value of the mixing angle. In this limit, the mass can be approximated as
\begin{equation}\label{eq:relaxmassgen}
m_\phi^2 \simeq 
\frac{v^2 \tilde M^2}{2f^2},
\end{equation} 
while the mixing angle reads
\begin{equation}\label{eq:sinmixgen1}
\sin \theta 
\simeq \frac {\tilde M^2}{2 \lambda_h f v}.
\end{equation}
Expressing $\tilde M$ through the relaxion mass, and using the lower bound on $f$ leads to a lower
bound on the mixing,
\begin{equation}\label{eq:genmixmin}
\sin \theta 
\simeq \frac{m_\phi^2 f}{\lambda_h v^3} > \frac{m_\phi^2 \Lambda_{\text{min}}}{\lambda_h v^3}\,.
\end{equation}
For the relaxion in such a minimum, and also for generic untuned Higgs portal models, the maximal
mixing is given by Eq.~(\ref{eq:genmixmax}).  All the $\sin\theta$ bounds derived in this section
are valid up to order-one factors and thus should not be taken as exact.

\paragraph{Combined constraints: the relaxion band}
As follows from the above discussion, for each mass $m_\phi$ there is a relaxion-specific lower and
upper bound on $\sin\theta$. The upper bound arises from the first minimum, see
Eq.~\eqref{eq:mixmax}, and always exceeds the upper bound for the relaxion in a generic minimum.
For $f_\mi=\Lambda_\mi=\tilde M_\ma = 1\tev$, the overall lower bound stems from the general minimum
for $m_\phi\leq 8\gev$ and from the first minimum otherwise.  This crossover causes a kink of the
lower bound.  The range of natural values of $\sin\theta$ for a given mass will appear as the
\textit{relaxion band} in the plots in the phenomenological analyses.

\begin{figure}[tb]
	\begin{center}
		\label{fig:RelaxionSpace_sth_lhp_m}
		\includegraphics[height=0.4\textheight]{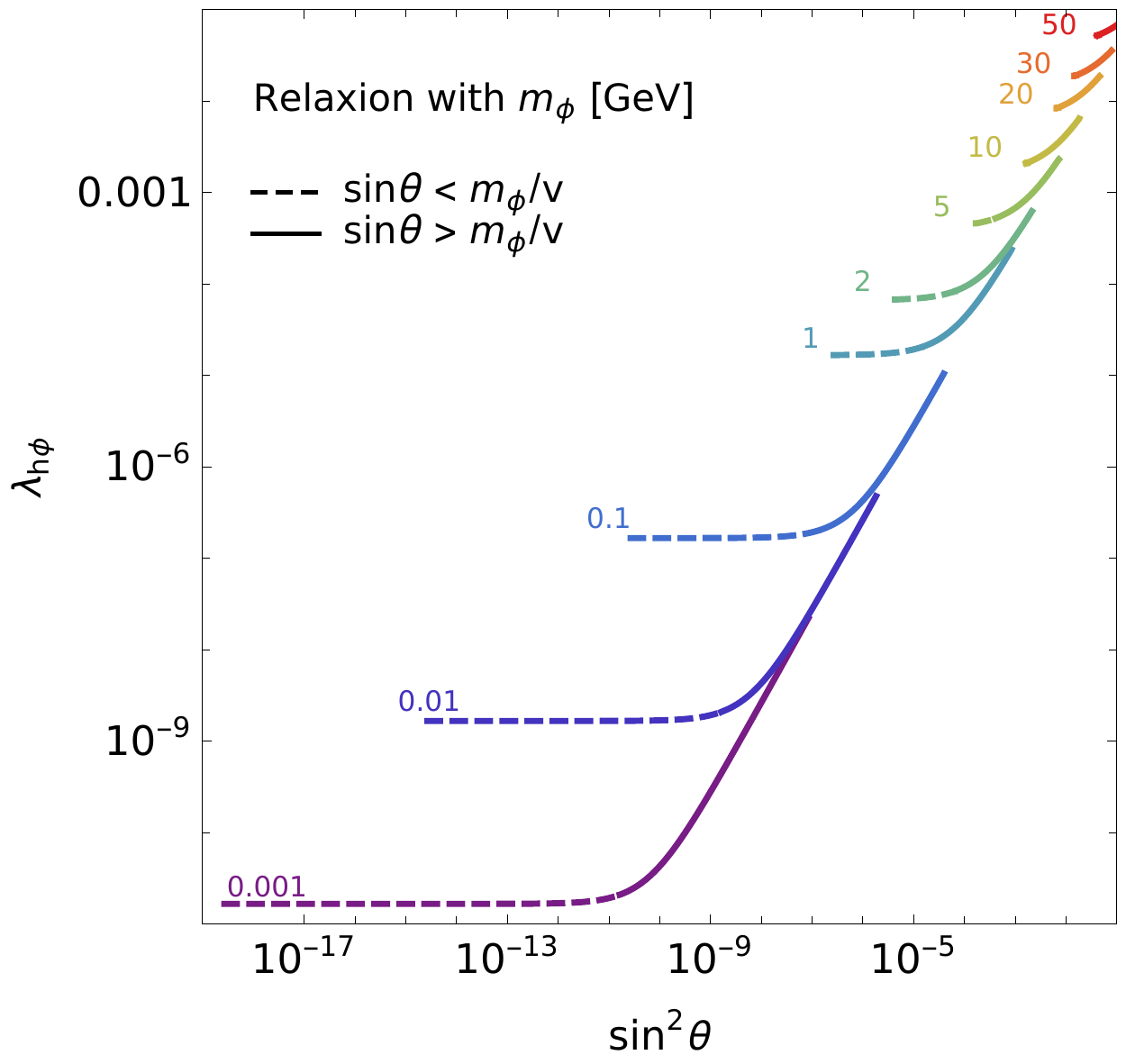}
		\caption{Natural relaxion parameter space in the $\sin^2\theta$-\lhp{} plane. Each
                  color shows one mass given in GeV. The dashed (solid) part of the lines
                  corresponds to $\sin^2\theta<(>)m_\phi^2/v^2$, \ie~where the mixing angle of the
                  renormalizable singlet is natural (unnatural). The plotted
                  $\lhp(m_\phi, \sin\theta)$ of the relaxion is defined in Eq.~\eqref{eq:lhpRel},
                  within the natural $\sin^2\theta$ range from Sec.~\ref{sec:RelaxionBand}. }
	\end{center}
\end{figure}

Fig.~\ref{fig:RelaxionSpace_sth_lhp_m} shows the lines in the $\sin^2\theta$-\lhp{} plane which
fulfill the relaxion relation for \lhp{} as a function of $m_\phi$ and $\sin\theta$ within the
$\sin^2\theta$ range bounded by naturalness of the first and generic minima. The dashed part of the
lines corresponds to $\sin^2\theta<m_\phi^2/v^2$, \ie~the naturalness limit of the renormalizable
singlet. The solid line segments represent $\sin^2\theta>m_\phi^2/v^2$, \ie~values that are
unnatural for the renormalizable singlet, but still natural for the relaxion.

\section{Collider bounds on (long-lived) scalar singlets}
\label{sec:collider-bounds-long}

\begin{figure}[tb]
	\begin{center}
		\label{fig:ctau}
		\includegraphics[height=0.3\textheight]{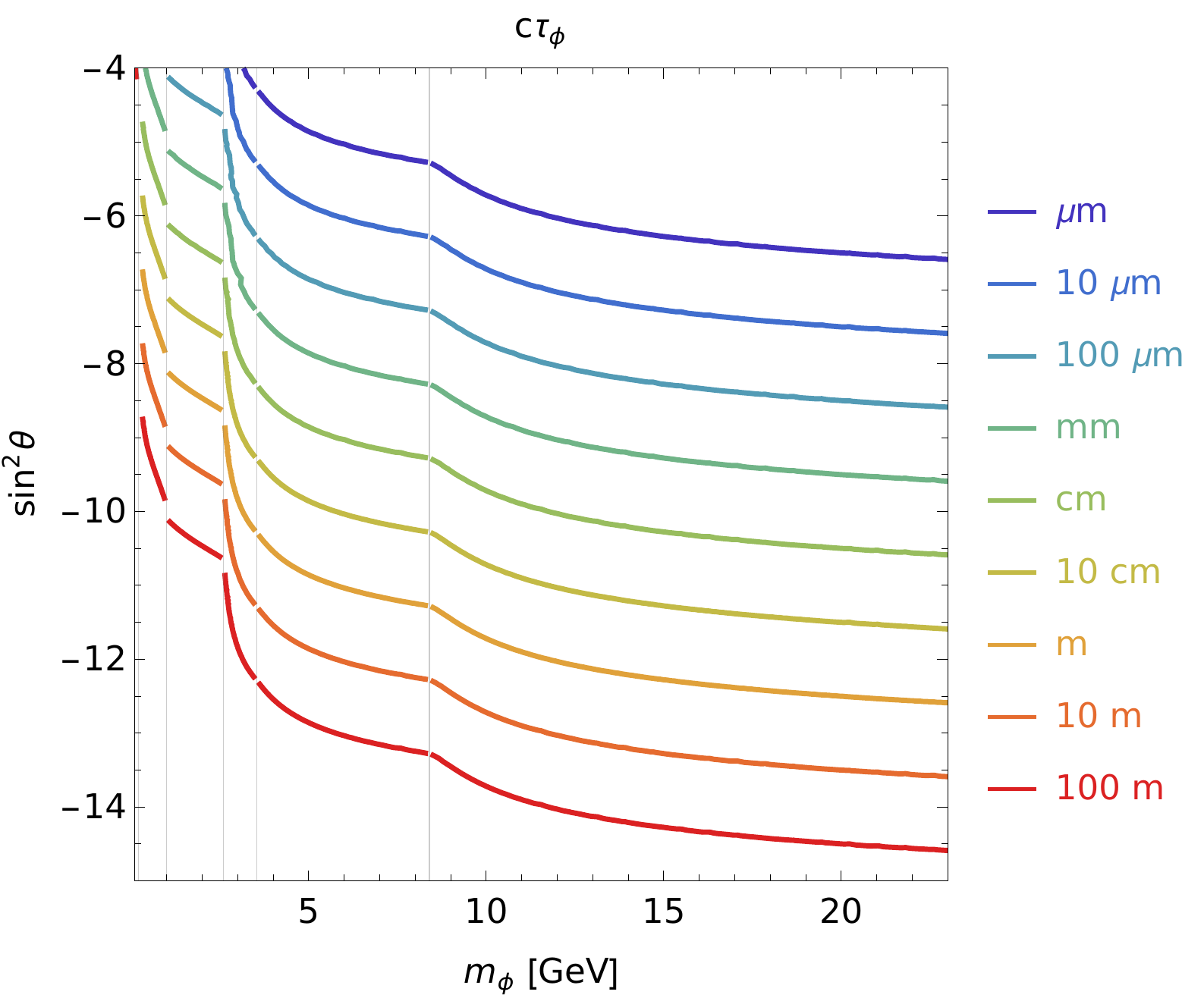}\\
		\caption{Decay length of the singlet~\cite{Flacke:2016szy} dependent on its mass $m_\phi$ and mixing angle
                  $\sin \theta$.}
	\end{center}
\end{figure}

We present bounds on scalar singlets for a broad range of their lifetime. This necessitates a
combination of various search strategies. Central to them is the lifetime which is shown in
Fig.~\ref{fig:ctau} for the relevant masses and mixing angles.
For
\begin{itemize}
\item short lifetimes, untagged Higgs decays into a pair of singlets lead to strong indirect bounds;
\item intermediate lifetimes, \ac{DV} searches and strategies based on timing information probe a
  broad range of the parameter space;
\item long lifetimes, the singlet escapes the detector and can account for invisible signatures.
\end{itemize}
We compare these bounds to the ones previously studied from direct searches in $Z$ decays and from
associated $Z\phi$ production. The presented bounds are based on singlet pair production via Higgs
decays ($h\to\phi\phi$). The production via singlet-Higgs mixing is negligible for the parameter
region considered here, for details see Appendix~\ref{sec:production-via-higgs}.  Our bounds apply
to the general singlet extension of Sec.~\ref{sec:general-singlet-extension}. We will point out
which regions of the displayed parameter space can also be realized by the relaxion.

\subsection{Fits of untagged and invisible Higgs decays}
\label{sec:untagged}
\ac{BSM} physics can modify the tagged Higgs branching ratios both by modifying the Higgs couplings
to \ac{SM} particles by $\kappa_x=c_x/c_x^\sm$, and by introducing new decay channels for the Higgs,
depleting the relative \ac{SM} contribution to the total decay
width~\cite{Bechtle:2014ewa,Frugiuele:2018coc}
\begin{align}
\br_{h\to x}&=\frac{\kappa_x^2 \Gamma^\sm_{h\to x}}{\sum_{y\in \text{SM}}\kappa_y^2\Gamma^\sm_{y}+\Gamma_\bsm }\approx \br^\sm_{h\rightarrow x}\leri{1-\br_\bsm}\,.\label{eq:brSM}
\end{align}
The \ac{BSM} particles produced in these Higgs decays can either decay visibly, or remain
invisible. While searches for Higgs decays with missing energy directly constrain the invisible
branching $\br_\inv$, these search results can also be used as a tagged category in a fit. In
contrast, the final states of the visible \ac{BSM} Higgs decays (\eg~light jets) are generally not
included in the list of tagged visible decays (such as $h\to \tau\tau,~bb,~VV,~...$, explicitly
displayed \eg{} in Tab.~1 of Ref.~\cite{Khachatryan:2014jba}). Hence they remain
\textit{untagged}\footnote{For the implications of the direct searches for $h\to\phi\phi$ further
  decaying promptly into four visible particles, \eg{}~$h\to4b$, $h\to bb\tau\tau$, see
  Refs.~\cite{Frugiuele:2018coc,deBlas:2018mhx,CEPCStudyGroup:2018ghi,CidVidal:2018eel}.}, and the
corresponding Higgs branching $\br_\unt$ is not determined by any specific search, but by the
uncertainties of the tagged channels. Therefore, global fits of the Higgs coupling modifiers
$\kappa_x$ to measured signal strengths $\mu_{if} = \sigma_i/\sigma_i^\sm \cdot \br_f/\br_f^\sm$
(tagged production cross sections times tagged branching ratios normalized to the \ac{SM}
prediction), together with searches for invisible Higgs decays, allow to constrain the Higgs decay
width into \ac{BSM} particles\footnote{The \ac{SM} contributions to $\Gamma_\inv$ and $\Gamma_\unt$
  are subtracted.}, $\Gamma_\bsm = \Gamma_\inv + \Gamma_\unt$.

\begin{figure}[ht!]
	\begin{center}
		\includegraphics[width=0.7\textwidth]{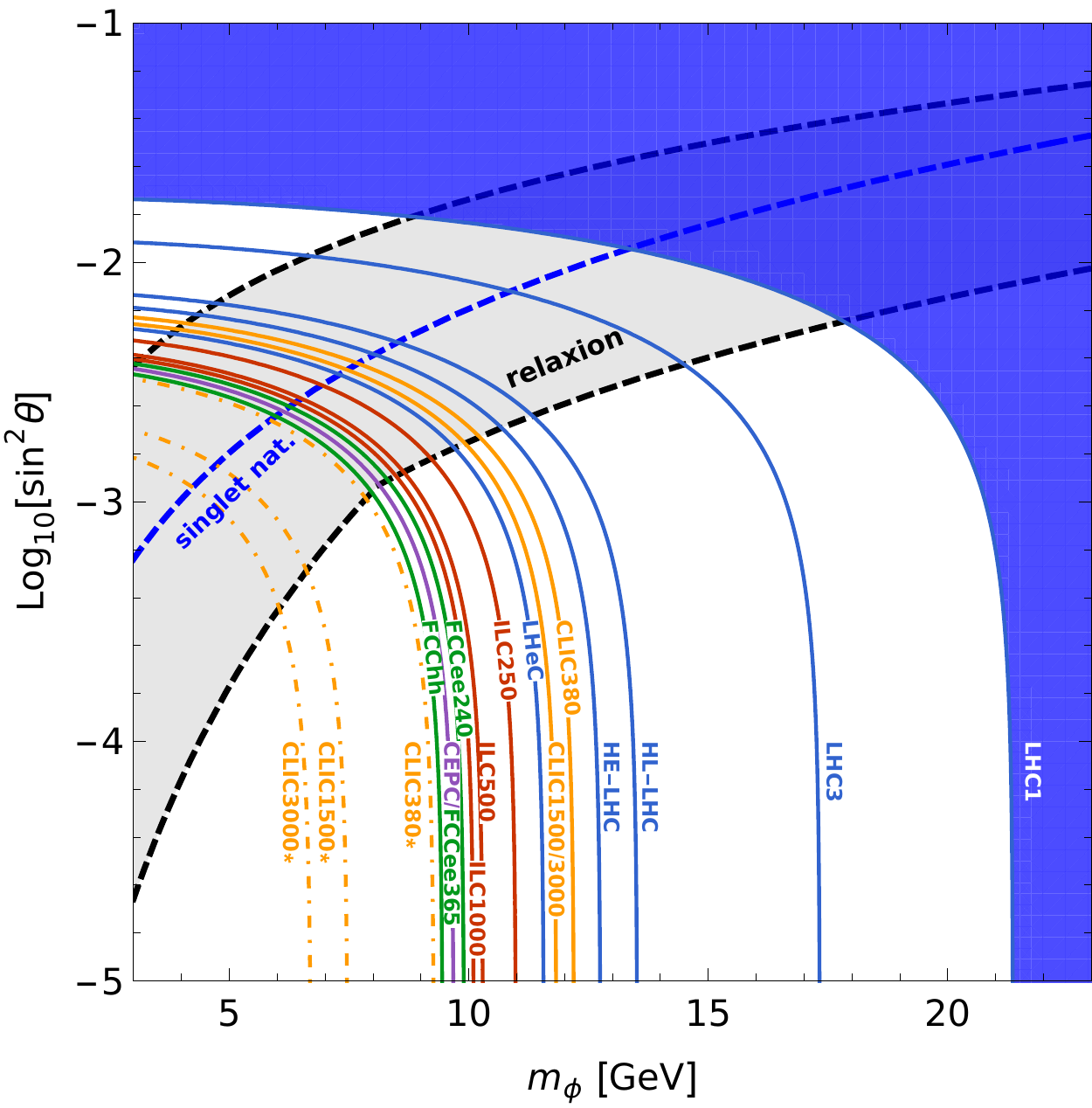}
		\caption{Existing and projected constraints on $\st$ and $\mphi$ from bounds on the
                  branching ratio of Higgs to untagged or BSM final states listed in
                  Tab.~\ref{tab:HGfit}.  The blue shaded area is already excluded. The limits from
                  FCCee at 365\,GeV and CEPC coincide (purple). CLIC at $3\tev$ does not improve the
                  CLIC limit at $1.5\tev$ (solid yellow).  The dash-dotted bounds for CLIC labelled
                  by a * indicate the sensitivity from the 1-parameter fit to $\br_\bsm$ valid in
                  the limit $\sin^2\theta\ll1$.  The dashed dark blue line represents the upper
                  naturalness bound $\sin\theta\leq m_\phi/m_h$ on the singlet from
                  Eq.~\eqref{eq:genmixmax}.  The gray band within the black dashed lines is the
                  natural relaxion range defined by Eqs.~\eqref{eq:mixmax},\,\eqref{eq:mixmin} and
                  \eqref{eq:genmixmin}.  }
		\label{fig:untagged}
	\end{center}
\end{figure}

The global Higgs fits performed in the scope of the European Strategy Update~\cite{deBlas:2019rxi}
present results for the future hadron colliders HL-LHC, LHeC, HE-LHC and FCChh, as well as for the
lepton colliders ILC, CLIC, CEPC and FCCee running at different energy stages.  Here we apply the
results from the so-called kappa-2 scenario that treats $\br_\inv$ and $\br_\unt$ as free parameters
for each collider individually. In addition, it has several independent $\kappa_x$ whereas in the
general singlet and the relaxion models there is only one overall $\kappa \equiv \cos\theta$, see
also Ref.~\cite{Frugiuele:2018coc}. Furthermore, in the region of intermediate and high
$\sin\theta\gtrsim 10^{-11}$, such that $c\tau_\phi(m_\phi, \sin\theta)\lesssim 1$\,m for
$m_\phi\geq 5\gev$ (see Fig.~\ref{fig:ctau}), all $\phi$s decay inside the detector, hence
$\br_\inv=0$, and fitting only two parameters, $\kappa$ and $\br_\unt$, would be sufficient. In the
opposite case of very small $\sin\theta$, the Higgs couplings to \ac{SM} particles become
\ac{SM}-like ($\kappa\simeq1$), and fitting only $\br_\inv$ would be enough. Hence, the
multi-parameter fits used in Tab.~\ref{tab:HGfit} and in Fig.~\ref{fig:untagged} give rise to
\textit{conservative} bounds on this actually more predictive model, defined by less parameters.  To
evaluate the gain in sensitivity by fitting only the needed parameters, we also include the
dedicated fit results performed for the CLIC stages~\cite{deBlas:2018mhx}, see the lower part of
Tab.~\ref{tab:HGfit}.

A combination of the ATLAS and CMS data collected in Run-1 results in a limit on
$\br_\bsm<20\%$~\cite{Bechtle:2014ewa} (which can applied be as a conservative bound on $\br_\unt$),
comparable to that of ATLAS alone in Run-2 of $\br_\unt<21\%$~\cite{Aad:2019mbh}. The strong result
of the Run-1 combination, despite the smaller summed luminosity, is due to the fit of only a global
$\kappa$ and $\br_\bsm$.  A Run-2 combination or a dedicated 2-parameter fit will be able to exclude
further parameter space based on the already existing data.

In Fig.~\ref{fig:untagged} we show the constraints on the $m_\phi$-$\sin^2\theta$ parameter plane of
the relaxion.  In addition, we show in gray the natural \textit{relaxion band}, whose upper and
lower $\sin\theta$ limits are discussed in Sec.~\ref{sec:relaxion-as-special}.  The experimental
limits and projections result from requiring
\begin{equation}
 \br_{h\to\phi\phi}(m_\phi,\,\sin^2\theta) = \frac{\Gamma_{h\to\phi\phi}}{(1-\sin^2\theta)\Gamma_{\text{tot}}^\sm + \,\Gamma_{h\to\phi\phi} } 
 \leq \br_\unt
\end{equation}
where the partial width $\Gamma_{h\to\phi\phi}\propto c_{h\phi\phi}^2$ is given in
Eq.~\eqref{eq:c_hphiphi_relaxion}, and the total Higgs width in the \ac{SM} is
$\Gamma_{\text{tot}}^\sm = 4.1\mev$~\cite{Heinemeyer:2013tqa}.  The contours form horizontal and
vertical asymptotes determined by the $\sin^2\theta$ and $m_\phi$ contributions to $c_{h\phi\phi}$,
respectively.  When neglecting the kinematical mass dependence of $\Gamma_{h\to\phi\phi}$ (for
$m_\phi\ll m_h/2$) and the \ac{BSM} contribution to the total width, the location of the asymptotes
for the relaxion can be approximated as
\begin{align}
 \left.\sin^2\theta\right\vert_{m_\phi\to0} 
    &\approx \frac{4v}{3}\sqrt{\frac{2\pi\,\br_\unt\,\Gamma_{\text{tot}}^\sm}{m_h^3}} 
    \simeq 0.038 \sqrt{\br_\unt}\label{eq:s2tAsymp}\,,\\
 \left.m_\phi\right\vert_{\st\to0} 
    &\approx \left(8\pi v^2\,m_h\,\Gamma_{\text{tot}}^\sm\,\br_\unt\right)^{1/4}
    \simeq 30\,\br_\unt^{1/4}\,.
\end{align}

The shaded blue area is already ruled out by Run-1 of the LHC, excluding natural mixing angles of
heavy relaxions above $m_\phi\gtrsim 18\gev$. As indicated in Tab.~\ref{tab:HGfit}, this Run-1 bound
is in fact on $\br_\bsm$, whereas for such large values of $\sin^2\theta$ all relaxions decay inside
the detector. Hence, a specific bound on $\br_\unt$ will exclude also lighter relaxions.  The
strongest bound will be reached by the FCChh, excluding $m_\phi\gtrsim 10\gev$ and
$\sin^2\theta\gtrsim 3\cdot 10^{-3}$.  As indicated by the dash-dotted yellow lines, the fit of only
$\br_\bsm$, assuming all $\kappa_x=1$, for CLIC leads to a significant improvement of the bound
compared to the multi-$\kappa$ fit at CLIC\footnote{Strictly, this fit is applicable only for
  vanishing $\sin^2\theta$, but in any case the exclusion contour of CLIC380 (CLIC3000) reaches only
  $\sin^2\theta\simeq3\cdot10^{-3}$ ($1.5\cdot 10^{-3}$) corresponding
  $\delta\kappa \equiv 1- \cos\theta \simeq 1.5\cdot 10^{-3}$ ($7.6\cdot10^{-4}$), \ie~just below
  the resolution of $\kappa$, see Tab.~\ref{tab:HGfit}. Hence we use this fit as an illustration of
  the gain of sensitivity in a suitable fit of such a predictive model}.  Dedicated fits for the
FCCee and FCChh could have the potential to close the high-mass relaxion window above few GeV.

The situation is different for the general singlet model where \lhp{} is a free parameter and
$\br_{h\to\phi\phi}$ varies with the choice of \lhp{}. For a larger value of \lhp{} than the one
predicted within the relaxion framework, the bounds from untagged Higgs decays can become even
stronger, whereas they get reduced to the $\sin^2\theta$ dependence in
Eq.~\eqref{eq:chphiphi_singlet_approx2} if \lhp{} is suppressed.  For a fixed \lhp{}, the bounds
only depend on $\sin^2\theta$ (up to the kinematical mass dependence), however, for small enough
masses, any fixed value of \lhp{} will eventually become unnatural, see Eq.~\eqref{eq:lhphimax}.
The naturalness upper bound on the mixing angle for the singlet is shown as the dashed blue line
(within the relaxion band).

In general, while the bounds on $\br_\bsm$ hold for arbitrary values of $\sin\theta$, the more
specific bounds on $\br_\unt$ are valid as long as the decay length is significantly smaller than
the detector size. Conversely, the bounds on $\br_\inv$ apply to decay lengths clearly exceeding the
detector size.
\begin{table}[tb]
	\begin{center}
		\begin{tabular}{|l|c|c||c|c|l|}
			\hline
			Collider &$\sqrt{s}$ [TeV]&$\lint$ [ab$^{-1}$] &$\textrm{BR}_{\unt}$ [\%] &$\delta\kappa$ [\%]&Ref.\\ \hline\hline
			LHC1	&$7, 8$		&0.022	&20 $\diamond $	&26	&\cite{Bechtle:2014ewa} Tab.~8,\,1\\
			LHC3 (S2)	&$13$		&0.3	&12.3 $\triangle$	&8.6	&\cite{Bechtle:2014ewa} Tab.~11\\ \hline
			HL-LHC&$14$		&6	&4	&0.99	&\multirow{3}{*}{\cite{deBlas:2019rxi} Tab.~28}\\
			HE-LHC (S2)&$27$		&15	&3.2	&0.99	& \\
			LHeC	&$1.3$		&1	&2.2	&0.99	& \\ \hline
			ILC250&$0.25$		&2	&1.8	&0.3	&\multirow{2}{*}{\cite{deBlas:2019rxi} Tab.~29}\\
			ILC500&$0.25,\,0.35,\,0.5$&2+0.2+4 &1.4 &0.24	&\\
			ILC1000&$0.25,\,0.35,\,0.5,\,1$&2+0.2+4+8 &1.3 &0.24	&\\\hline
			CEPC	&$M_Z, 2M_W,0.24$&16+2.6+5.6&1.1&0.19	&\cite{deBlas:2019rxi} Tab.~29\\\hline
			FCCee240&$0.24$	    &5	      &1.2	&0.21	&\multirow{3}{*}{\cite{deBlas:2019rxi} Tab.~29}\\
			FCCee365&$0.365$	&1.7	&1.1	&0.18	&\\ 
			FCCee/eh/hh&$100$	&30	      &1	&0.17	&\\ \hline
			TeraZ	  &$M_Z$  	&$N_Z=10^{12}$	&	&	&\\ \hline
            CLIC380&$0.38$	&1	&2.7	&0.5	&\multirow{3}{*}{\cite{deBlas:2019rxi} Tab.~29}\\
			CLIC1500&$1.5$ 	&2.5	&2.4	&0.39	&\\
			CLIC3000&$3$	&5	    &2.4	&0.38	&\\ \hline
			CLIC380&$0.38$	&1	&0.92 $\star$	&0.58 $\diamond$	&\multirow{3}{*}{\cite{deBlas:2018mhx} Tab.~6}\\
			CLIC1500&$1.5$ 	&2.5	&0.39 $\star$	&0.57 $\diamond$	&\\
			CLIC3000&$3$	&5	&0.26 $\star$	&0.57 $\diamond$	&\\ \hline
		\end{tabular}
		\caption{ Upper bounds on $\br(h\to\unt)$ at 95\%\,CL from global fits of Higgs
                  signal strengths for different colliders.  $\diamond$: 2-parameter fit of $\kappa$
                  and $\br_\bsm$; $\triangle$: fit of multiple $\kappa_x$ and $\br_\bsm$; $\star$:
                  1-parameter fit of $\br_\bsm$ (applicable to low $\sin\theta$ because
                  $\kappa\equiv1$); if not labeled, then multi-$\kappa$ fit of $\br_\unt$.
                  $\br_\bsm$ can be interpreted as a \textit{conservative} $\br_\unt$ bound.  The
                  LHC Run-3 bound at approximately 95\% CL was obtained by multiplying the 68\% CL
                  bound by 1.3, the ratio of the quantiles of a $\chi^2$ distribution with 7
                  parameters. $\delta\kappa$ denotes the 68\% CL uncertainty of the modifier of the
                  most precisely determined Higgs coupling, \ie~$\delta\kappa_Z$ (except for the
                  high-energy stages of CLIC where $\delta\kappa_W$ is smaller).  }
		\label{tab:HGfit}
	\end{center}
\end{table}

\subsection{Displaced jets}
\label{sec:displaced}
\begin{figure}[tb]
	\begin{center}
		\subfloat[ATLAS  $\sqrt{s}=13$~TeV with $36\ifb$.]{\includegraphics[trim={0 2.5cm 0 2.5cm},clip,scale=0.32]{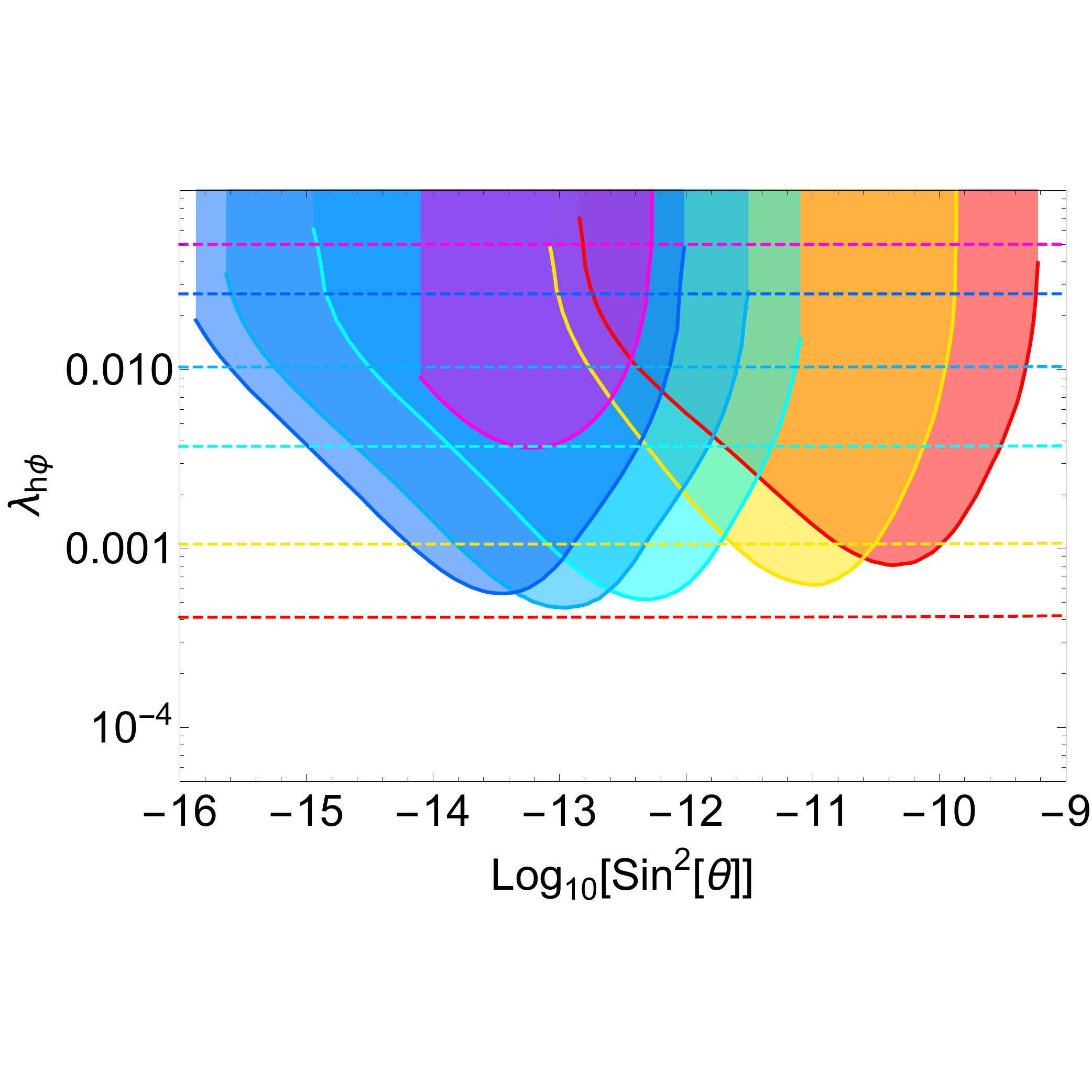}}
		\subfloat[FCC-ee $\sqrt{s}=240$~GeV with $5\iab$.]{\includegraphics[trim={0 2.5cm 0 2.5cm},clip,scale=0.32]{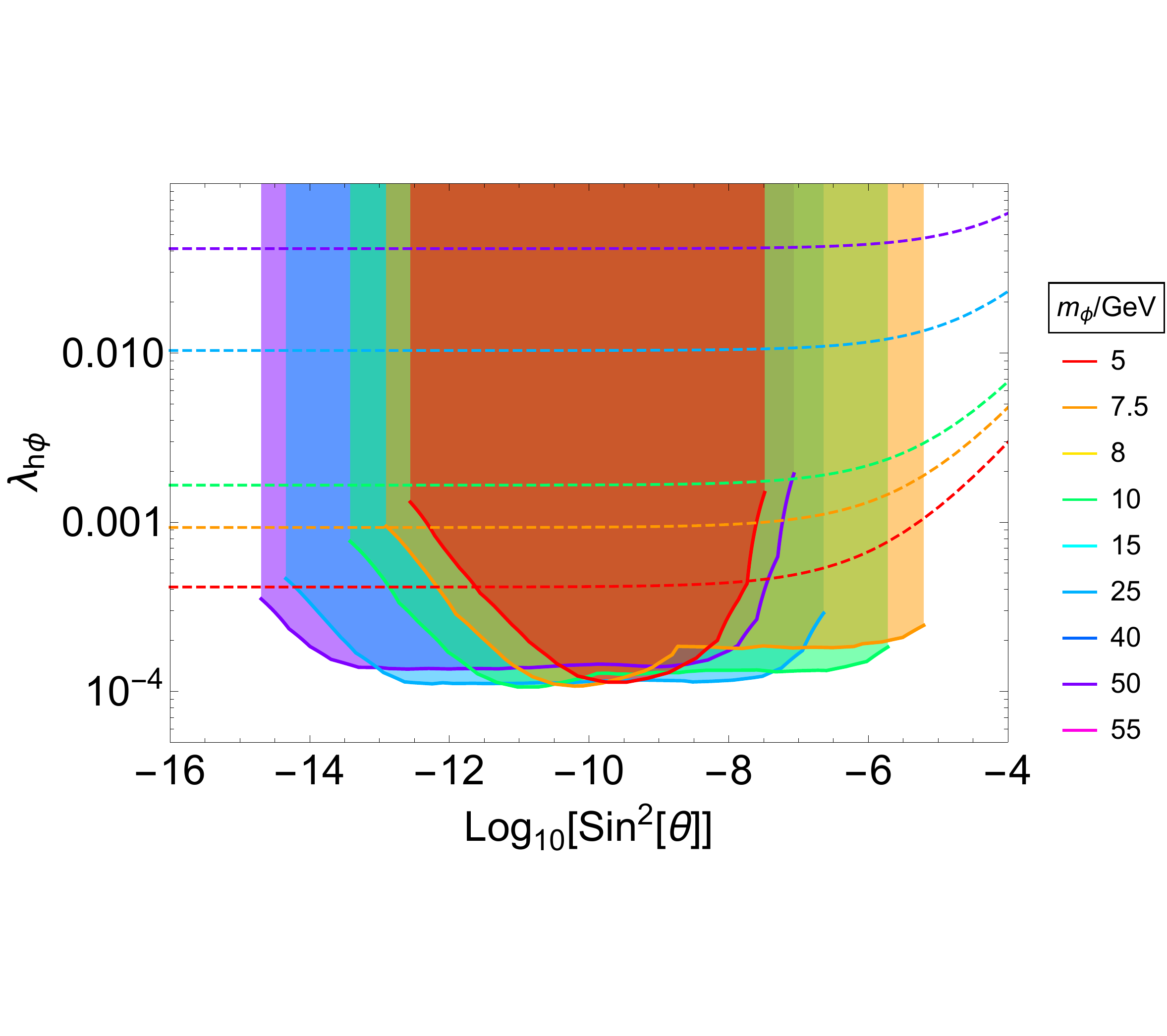}}
		\caption{Bounds on $\lambda_{h\phi}$ and $\st$ for various singlet masses arising
                  from searches for displaced jets in Higgs decays. The dashed lines show the upper
                  naturalness limit
                  \mbox{$\lambda_{h\phi}^{\rm{max}}=m_\phi^2/v^2 + 4\pi m_{\phi} \sint/v$}. }
		\label{fig:DV}
	\end{center}
\end{figure}

The singlet can be detected in searches for Higgs decays into displaced jets if it is sufficiently
long-lived, but still decays in the detector. ATLAS
searches~\cite{Aaboud:2018aqj,Aaboud:2019opc,Aad:2019xav} and FCC-ee
projections~\cite{Alipour-Fard:2018lsf} provide upper bounds on the branching ratio
$\text{BR}_{h\rightarrow \phi\phi}$ as a function of the proper decay length $c\tau_\phi$ for a few
singlet masses\footnote{For Higgs decays into complex singlets at the LHeC, see the recent
  Ref.~\cite{Cheung:2020ndx}.}. We transform them into upper limits on $\lambda_{h\phi}$ as a
function of $\sin^2\theta$, for the corresponding mass points given in the analyses, shown in
Fig.~\ref{fig:DV}. The dashed lines show the upper limit on $\lambda_{h\phi}$ from naturalness, see
Eq.~\eqref{eq:lhphimax}.  While for $m_\phi=5\gev$ the ATLAS searches do not constrain any natural
parameters of the singlet model, for higher masses the searches already probe parts of the natural
parameter region.  In contrast, FCC-ee will access natural parameter space for all masses. The
displayed FCC-ee bounds show the combination of the two analysis strategies from
Ref.~\cite{Alipour-Fard:2018lsf}, and therefore span a larger range of $\sin^2\theta$.  The CLIC
sensitivity to a long-lived scalar singlet via displaced vertex searches was studied in
Ref.~\cite{Strategy:2019vxc} and is included in our overview plot in Fig.~\ref{fig:overview}. The
comparison shows that CLIC and FCCee provide a comparable sensitivity.

\subsection{Delayed jets}
\label{sec:timing}
      \begin{figure}[t!]
	\begin{center}
		\subfloat[HL-LHC, $\sqrt{s}=14$~TeV.]{\includegraphics[height=180pt]{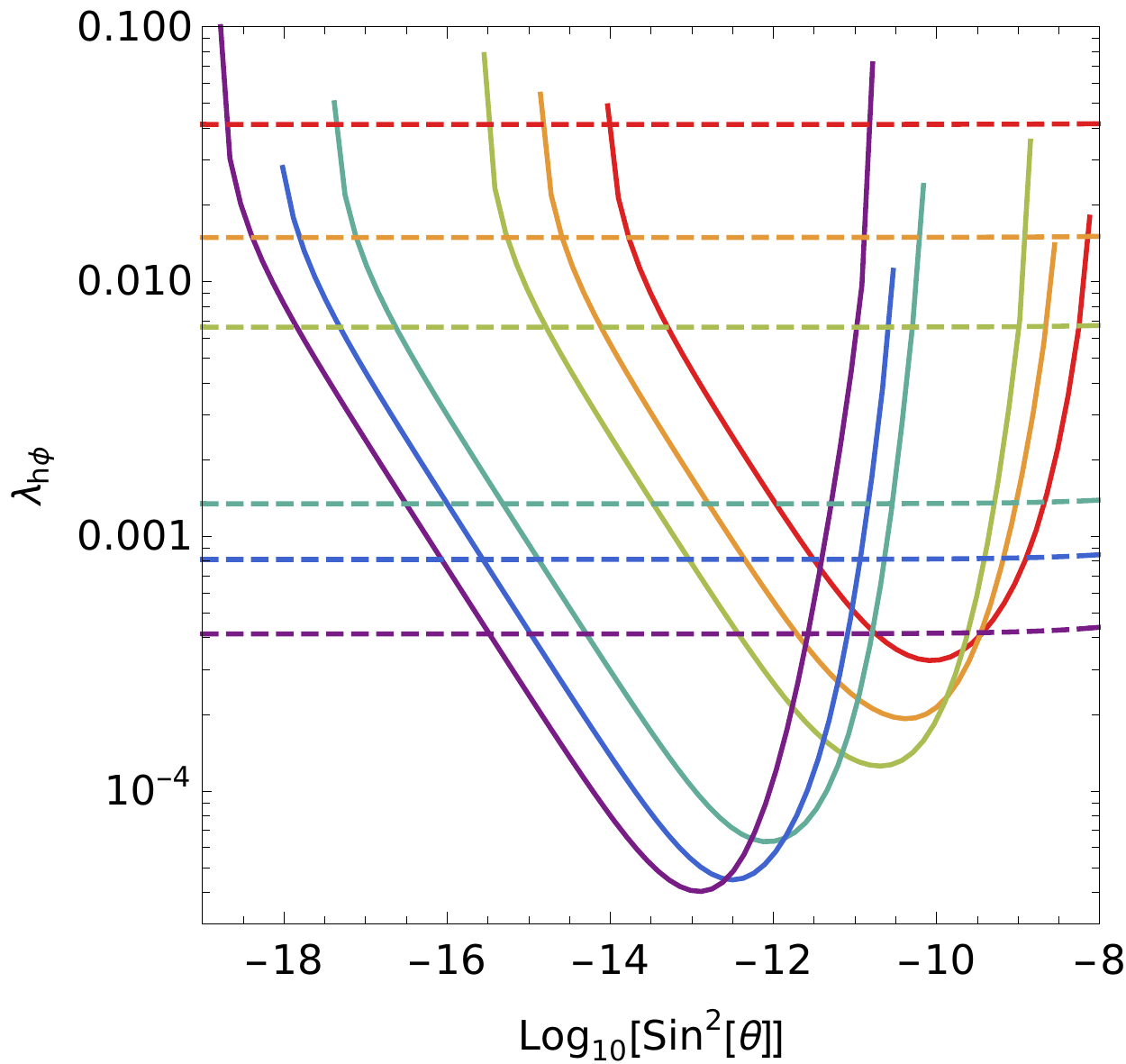}}	
		\subfloat[FCC-ee, $\sqrt{s}=240$~GeV.]{\includegraphics[height=180pt]{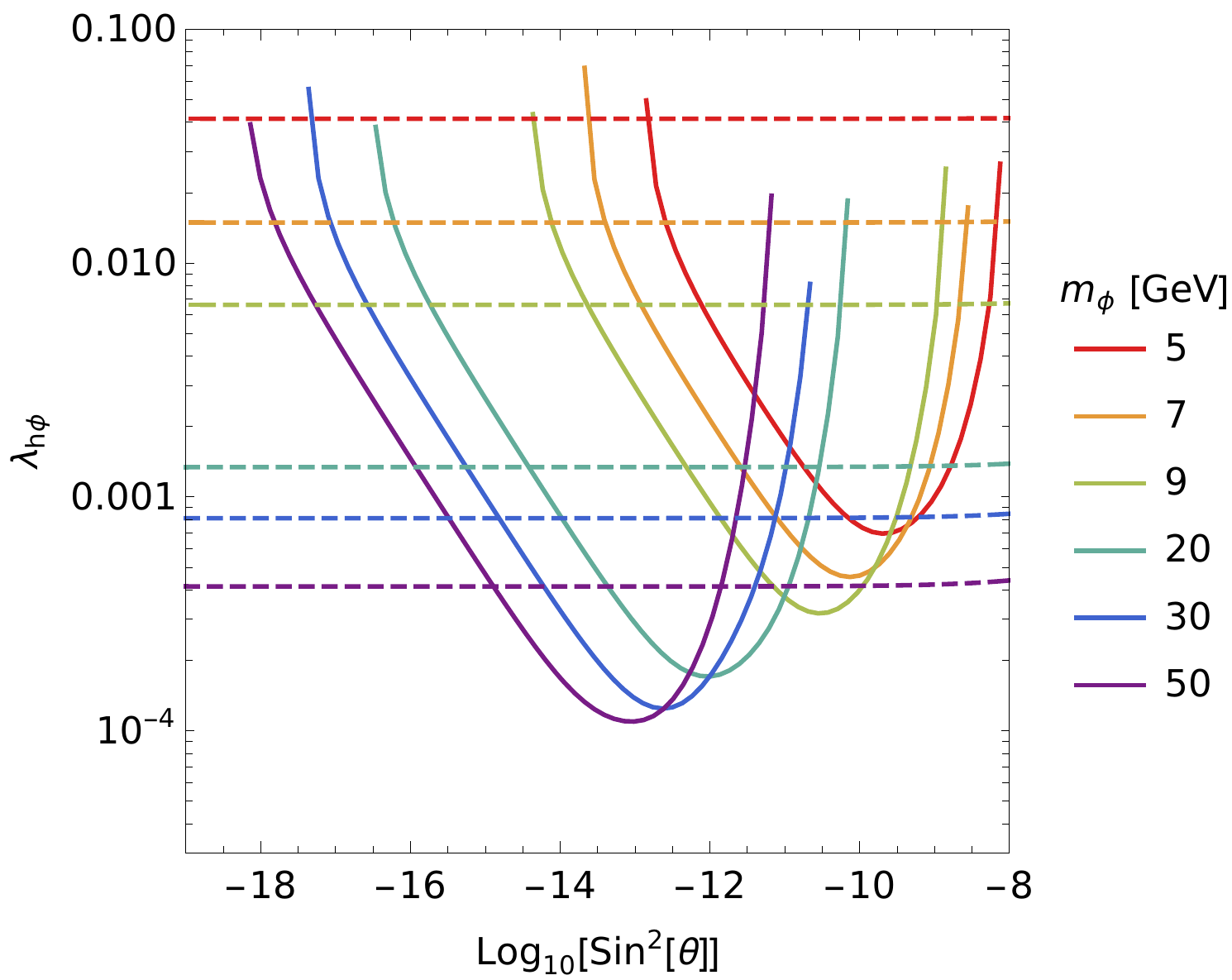}}
		\caption{Bounds on $\lambda_{h\phi}$ and $\st$ for various singlet masses arising
                  from searches for delayed jets in Higgs decays. The dashed lines show the upper
                  naturalness limit $\lhp^\textrm{max}$ of for each mass.
                  }
		\label{fig:Delayed}
	\end{center}
\end{figure}
A powerful strategy to search for long-lived particles was recently presented in
Ref.~\cite{Liu:2018wte}, allowing to detect displaced vertices in the CMS tracker\footnote{The new
  proposal in Ref.~\cite{Liu:2020vur} evaluates the sensitivity of the High Granularity Calorimeter
  of the CMS detector upgrade to the same type of $h\to\phi\phi$ decays. While the conservative
  estimate yields bounds comparable to the timing bounds of Ref.~\cite{Liu:2018wte}, only the
  analysis assuming a displaced track trigger could improve them. }.  This proposal utilizes the
timing detector layer, to be installed at the \ac{HL}-\ac{LHC}~\cite{timingDetectors}, to identify
secondary vertices by the delayed arrival, $\Delta t$, of the light decay products, compared to the
arrival time expected for a directly travelling SM particle. An~\ac{ISR} jet is used to time-stamp
the collision.  Ref.~\cite{Liu:2018wte} provides the bounds for the benchmark scalar masses of
$m_\phi=10\gev$ and $50\gev$ at the \ac{HL}-\ac{LHC}.  In order to determine the mass dependence of
the experimental reach, we simulate Higgs events at the LHC and FCC-ee, using
MadGraph5~\cite{Alwall:2014hca} at \ac{LO}, where the Higgs decays by \mbox{$h\to\phi\phi$}, and
each scalar decays through \mbox{$\phi\to jj$}.  Subsequently, we implement the search strategy
presented in~\cite{Liu:2018wte}, reproduce its results, and apply it to the additional mass
points. For the FCC-ee, we assume a (hypothetical) timing detector comparable to the one planned for
the \ac{HL}-\ac{LHC}. The detection efficiency is mostly affected by demanding a long time delay of
the jet produced in the singlet decay, related to the singlet's path through the detector, along
with requiring the singlet to decay between the inner tracker and the timing layer.  Hence, the
selection criteria for this search are mainly geometrical.  Therefore, for each event kinematics and
for each jet $j$ in the event, we find the range of lab frame singlet decay lengths $l_\phi$ for
which an event will be accepted.  Since the detection of a single delayed jet is sufficient, each
event is then weighed by the event efficiency
\mbox{$\epsilon_\text{event}=1-(1-w_1)(1-w_2)(1-w_3)(1-w_4)$}, where $w_j$ is the probability of
$\phi$ to decay within the allowed region, which is calculated from an exponential distribution
\begin{equation}\label{eq:w}
w_j=\frac{1}{c\tau_\phi\gamma_\phi\beta_\phi}\int_{l_{\phi,j} \text{allowed}}\exp\leri{-\frac{l}{c\tau_\phi\gamma_\phi\beta_\phi}} \text{d}l\,.
\end{equation}
More details on the calculation, as well as on the resulting efficiencies and the expected upper
limits on \mbox{$\text{BR}_{h\rightarrow \phi\phi}$} as a function of $c\tau_\phi$ can be found in
Appendix~\ref{appendix:timing}.

The interpretation of these bounds in terms of the singlet parameters $\lambda_{h\phi}$ and
$\sin^2\theta$ is shown in Fig.~\ref{fig:Delayed}. While the HL-LHC probes natural values of $\lhp$
for $m_\phi>5\gev$, at the FCC-ee this is the case only for slightly higher masses.  As this
analysis has almost zero background in the signal region of $\Delta t > 1\,\text{ns}$ (for details
see Ref.~\cite{Liu:2018wte}), its sensitivity is determined by the number of Higgses.  Therefore, the HL-LHC appears to perform better than the
FCCee. Since it is the hadronic environment at the HL-LHC that necessitates this restrictive cut on
$\Delta t$, the FCCee can allow for a looser cut, and the limit presented here based on the HL-LHC
cut is conservative.

\subsection{Searches for invisible Higgs decays}
\label{sec:invisible}

If the proper decay length of the scalar is larger than, or comparable to, the size of the detector,
the scalar may give rise to missing energy.  Global Higgs coupling fits set strong bounds on
$\text{BR}_{h\to\inv}$~\cite{deBlas:2019rxi}. These can be interpreted as bounds on \lhp{} in the
limit of vanishing $\sin\theta$, \ie~infinite lifetime.  To investigate the region of intermediate
lifetimes where only a fraction of the scalars escape the detector, we need to make use of direct
searches for invisible Higgs decays.  To take this fraction into account, we recast the analysis by
CMS and the studies for the HL-LHC and FCCee listed in Tab.~\ref{tab:Invisible} to constrain the
appropriate region of the singlet parameter space.  The bounds given by these searches need to be
weakened by a factor $r$, accounting for the cases where both scalars decay outside the detector.
The rescaling factor $r$ is obtained by
\begin{align}
r=\frac1N\sum^N_{i=1}\exp\leri{-\frac{m_\phi}{c\tau_\phi }\leri{\frac{L_{i_1}}{p_{i_1}}+\frac{L_{i_2}}{p_{i_2}}}}\,, \label{eq:r_calc}
\end{align}
where the sum runs over all $h\rightarrow \phi \phi$ events passing the selection criteria when an
infinite decay length is assumed, $p$ is the momentum of each scalar, $L$ is the distance the scalar
travels inside the detector, and the indices \(\{1,2\}\) mark the two scalars produced in the Higgs
decay. A conservative estimate of the rescaled bound can be given by minimizing $r$ for each
search. For LHC searches, which require a large missing $p_T$, this can be approximated by
$r^\text{consv.}_\text{LHC}\approx\text{exp}\leri{-\frac{4L_Tm_\phi}{c\tau p_T^\text{miss} }}$ where
$L_T$ is the transverse detector size and $p_T^\text{miss}$ is the minimally required missing
transverse momentum.  For lepton colliders, such as the FCC-ee with a lower $\sqrt{s}=240$ GeV, a
better approximation is given by setting the energy of each scalar to $m_h/2$, as the Higgs is
produced at low momentum, \ie{}~
\mbox{$r^\text{consv.}_\text{FCC-ee}\approx\text{exp}\Big(-\frac{4Lm_\phi}{c\tau_\phi\sqrt{m_h^2-4m_\phi^2}}
  \Big)$}.

\begin{figure}[tb]
  \centering
  \includegraphics[width=0.7\textwidth]{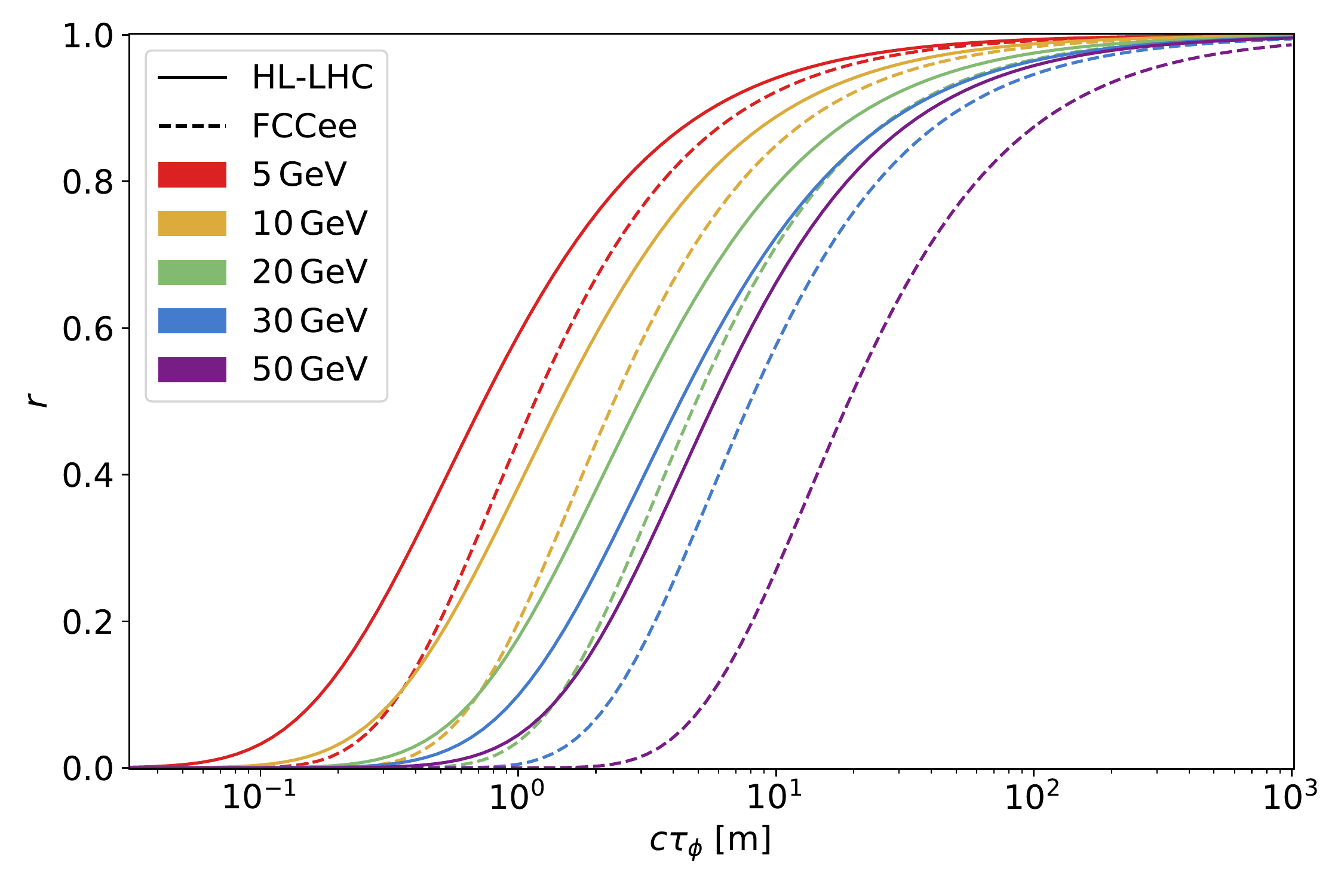}
  \caption{The rescaling factor $r$ defined in Eq.~\eqref{eq:r_calc} as a function of $c\tau_\phi$
    for the HL-LHC and FCCee. The dependence at the LHC is comparable to the HL-LHC.  The larger
    $r$, the more singlets escape the detector before decaying.  }
  \label{fig:r_calc}
\end{figure}

For a more precise estimate of the bounds, we determine $r$ for each search in
Tab.~\ref{tab:Invisible}. We use MadGraph5~\cite{Alwall:2014hca} to simulate the leading signal
process in each search at \ac{LO}. We then apply their selection cuts, and obtain the
\(\leri{\frac{L_{i_1}}{p_{i_1}}+\frac{L_{i_2}}{p_{i_2}}}\) distribution for each signal mass, and
subsequently obtain $r$ following Eq.~\eqref{eq:r_calc}. The signal processes and selection cuts
applied are summarized in Tab.~\ref{tab:simulationInvisible}. The resulting $r$ for the HL-LHC and
FCC-ee is shown in Fig.~\ref{fig:r_calc} as a function of $c\tau_\phi$. For a given $m_\phi$ and
$c\tau_\phi$, $r$ is larger for the HL-LHC because the $L/p$ distributions peak at lower values than
for the FCC-ee.

\begin{figure}[tb!]
		\begin{center}
                  \includegraphics[trim={0 4cm 0
                    4cm},clip,height=0.28\textheight]{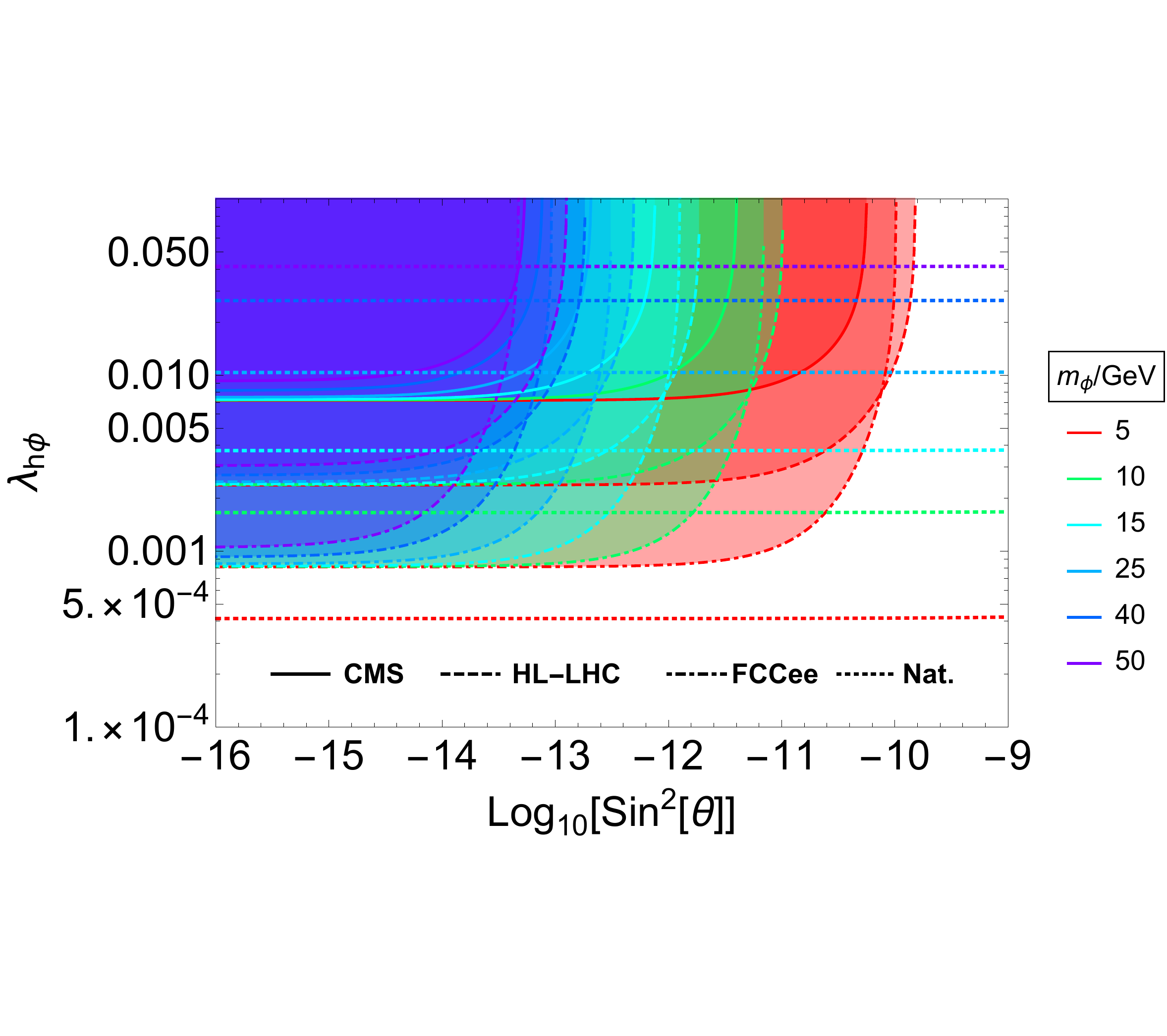}
                  \caption{Bounds on $\lambda_{h\phi}$ and $\st$ for various singlet masses arising
                    from searches for invisible Higgs decays. The dotted lines show the upper
                    naturalness limit $\lhp^\textrm{max}$ of for each mass.}
		\label{fig:INV}
	\end{center}
\end{figure}

The CMS bounds as well as the HL-LHC and FCCee projections on \lhp{} and $\sin^2\theta$ are shown
for different values of $m_\phi$ in Fig.~\ref{fig:INV}.  In general, each contour has a horizontal
and a vertical asymptote, driven by the limit on $\br_{h\to\phi\phi}$ and by the lifetime,
respectively.  As a consequence, the horizontal asymptotes are hardly mass dependent (apart from
$m_\phi=50\gev$ which is near the decay threshold), whereas the reach in $\sin^2\theta$ is larger
for low $m_\phi$ -- owing to the longer lifetime.  While for $m_\phi=5\gev$ no natural parameter
space is probed, for $m_\phi=10~(15)\gev$ only FCCee (and HL-LHC) access the natural parameter
space, and for higher masses this is also achieved in the present CMS analysis.

For the FCChh, the vast amount of produced Higgses can result in a very strong upper limit on the
invisible branching ratio.  Ref.~\cite{L.Borgonovi:2642471} reports for a luminosity of $30\iab$ an
expected sensitivity of a direct search to \mbox{$\br_{h\to\inv} \lesssim 3\cdot 10^{-4}$},
\ie~similar to the result from a global fit of
\mbox{$\br_{h\to\inv} \leq 2.4\cdot 10^{-4}$}~\cite{deBlas:2019rxi}.  The asymptotic limit on \lhp{}
for vanishing $\sin^2\theta$ can be approximated as
\begin{align}
  \lhp = \frac{2}{v} \sqrt{\frac{2\pi m_h \Gamma_{\text{tot}}^\sm\,\br_\inv}{\sqrt{1- \frac{4m_\phi^2}{m_h^2}}}}\,.\label{eq:lhp_inv}
\end{align}
This translates into the asymptotic bound on \lhp{} for $m_\phi = 5\gev$ ($50\gev$) of
$\lhp \leq 2.3\cdot10^{-4}$ ($2.9\cdot 10^{-4}$) using the fit result, hence stronger than the limit
of the direct searches for $h\to\inv$ at the FCCee, and probing natural values of \lhp{} throughout
this mass range. The approximate FCChh bounds are included in Figs.~\ref{fig:portalAll} and
\ref{fig:overview} up to values of $\sin^2\theta$ for which all singlets can be safely assumed to
decay outside the detector.

\begin{table}[tb!]
	\begin{center}
		\begin{tabular}{|l|c|c||c|c|}
			\hline
			Collider &$\sqrt{s}$ [TeV]&$\lint$ [ab$^{-1}$] & $\textrm{BR}_\text{inv}$ [\%] &Ref.\\ \hline\hline
			LHC2+LHC1	&$7,8,13$&0.005, 0.020, 0.036	&19 &\cite{Sirunyan:2018owy}\\
			HL-LHC &$14$	&3	& 2.5 &\cite{Cepeda:2019klc}\\
			FCCee240&$0.24$	&5	& 0.3 &\cite{Azzi:2012yn,Benedikt:2651299}\\ \hline
		\end{tabular}
		\caption{Analyses of invisible Higgs decays recast in this work to constrain the
                  scalar singlet.}
		\label{tab:Invisible}
\vspace{2em}
		\begin{tabular}{|c|c|c|c|c|}
			\hline
			Collider &$\sqrt{s}$ [TeV]& process & selections &Ref.\\ \hline\hline
		\multirow{2}{*}{LHC2}&\multirow{2}{*}{$13$}& \multirow{2}{*}{VBF} & $p_T^j\geq 80(40)\gev$ &	\multirow{2}{*}{\cite{Sirunyan:2018owy}}\\ 
		& & & $|{\Delta \eta_{jj}}|\geq 1$ &  \\
		\centering{+}& & & $|{\Delta \phi_{jj}}|\leq 1.5$\,rad &  \\
		& & & ${\eta_{j1}\eta_{j2}}\leq 0$ &  \\
		LHC1& & & $\text{min}\left|\Delta \phi\leri{p_T^j,p_T^\text{miss}}\right|\geq 0.5$\,rad &  \\
		& & & $E_T^{\text{miss}}\geq 250\gev$ &  \\
		& & & $m_{jj}\geq 200\gev$ &  \\ \hline
		\multirow{2}{*}{HL-LHC}&\multirow{2}{*}{$14$}& \multirow{2}{*}{VBF} & $p_T^j\geq 80(40)\gev$ &	\multirow{2}{*}{\cite{Cepeda:2019klc}}\\ 
		& & & $|{\Delta \eta_{jj}}|\geq 4$ &  \\
		& & & $|{\Delta \phi_{jj}}|\leq 1.8$\,rad &  \\
		& & & $\text{min}\left|\Delta \phi\leri{p_T^j,p_T^\text{miss}}\right|\geq 0.5$\,rad &\\
		& & & $E_T^{\text{miss}}\geq 190\gev$ &  \\
		& & & $m_{jj}\geq 2500\gev$ &  \\ \hline
			\multirow{2}{*}{FCCee240}&\multirow{2}{*}{$0.24$}& Higgs-strahlung: & $p_T^\ell,\,p_T^{\ell\ell}\geq 10\gev$ &	\multirow{2}{*}{\cite{Azzi:2012yn,Benedikt:2651299}}\\
		&  & $e^+e^-\to Z h $ & $p_L^{\ell\ell}\leq 50\gev$ & \\ 
			&  &$Z\to \ell^+\ell^-,~h\to\phi\phi$ &  $|m_{\ell\ell} - M_Z|\leq 4\gev$& \\ \hline
		\end{tabular}
		\caption{Signal processes and selection cuts applied in the calculation of the
                  fraction $r$ of invisible signal events. The $p_T^j$ cuts refer to the leading (subleading) jet.}
		\label{tab:simulationInvisible}
	\end{center}
\end{table}

\section{Overview}
\label{sec:overview}
Having presented details about each search strategy in the previous section, here we compile them
for comparison, to highlight their complementarity and to evaluate the probed parameter regions,
both for the general singlet and the relaxion.

\begin{figure}[tb]
	\begin{center}
		\includegraphics[trim={0 5cm 0 5cm},clip,scale=0.31]{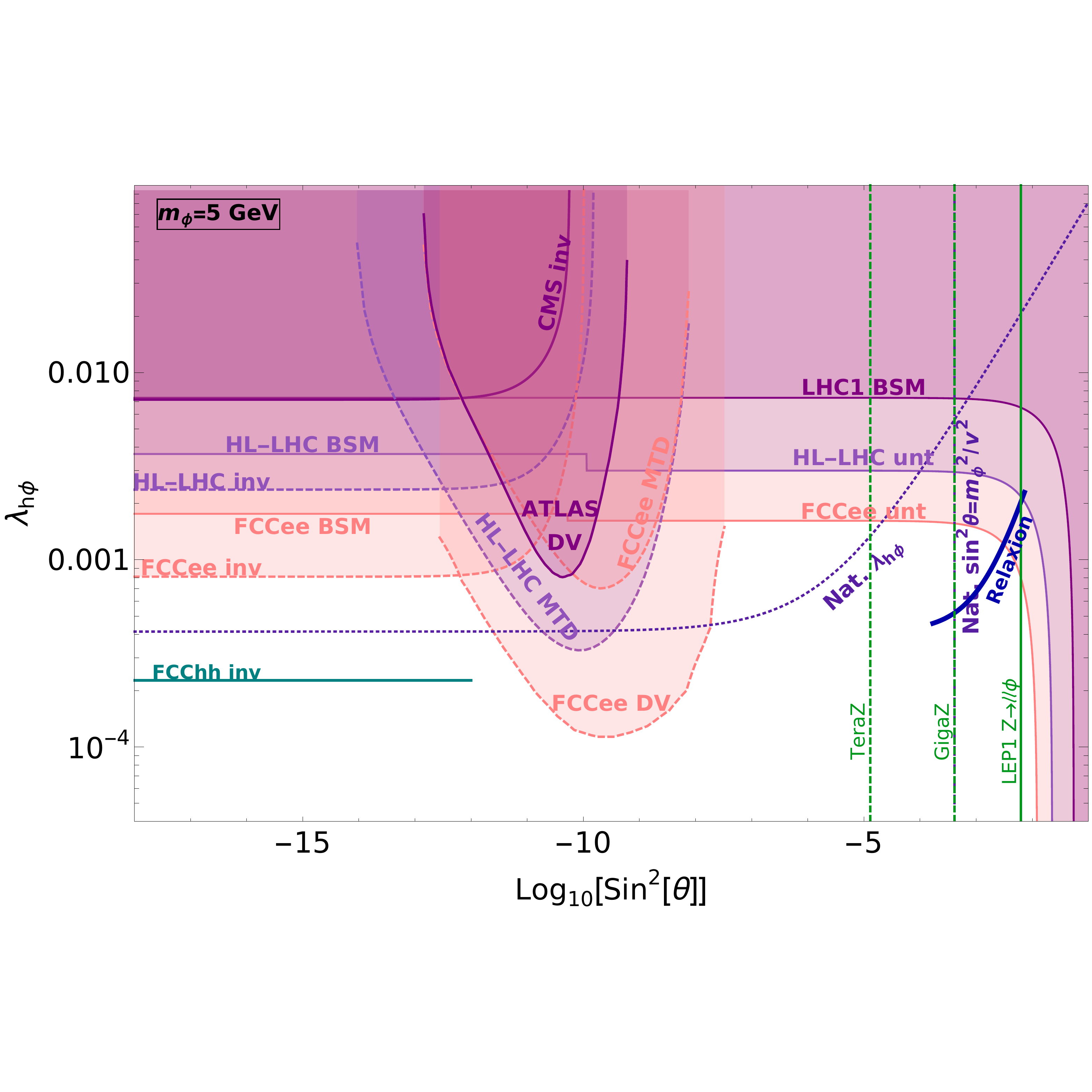}\\
		\includegraphics[trim={0 5cm 0 5cm},clip,scale=0.31]{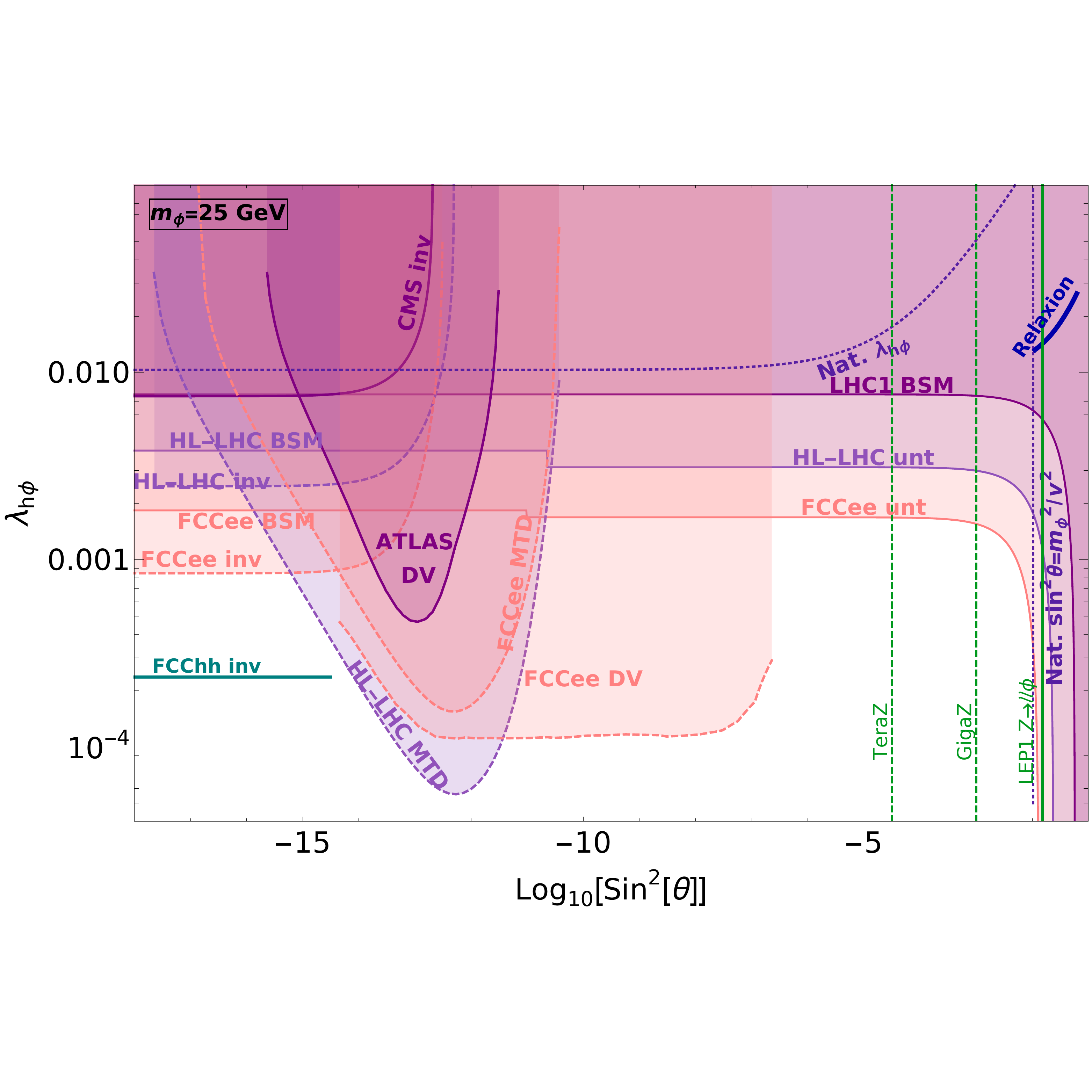}\\
		\includegraphics[trim={0 5cm 0 5cm},clip,scale=0.31]{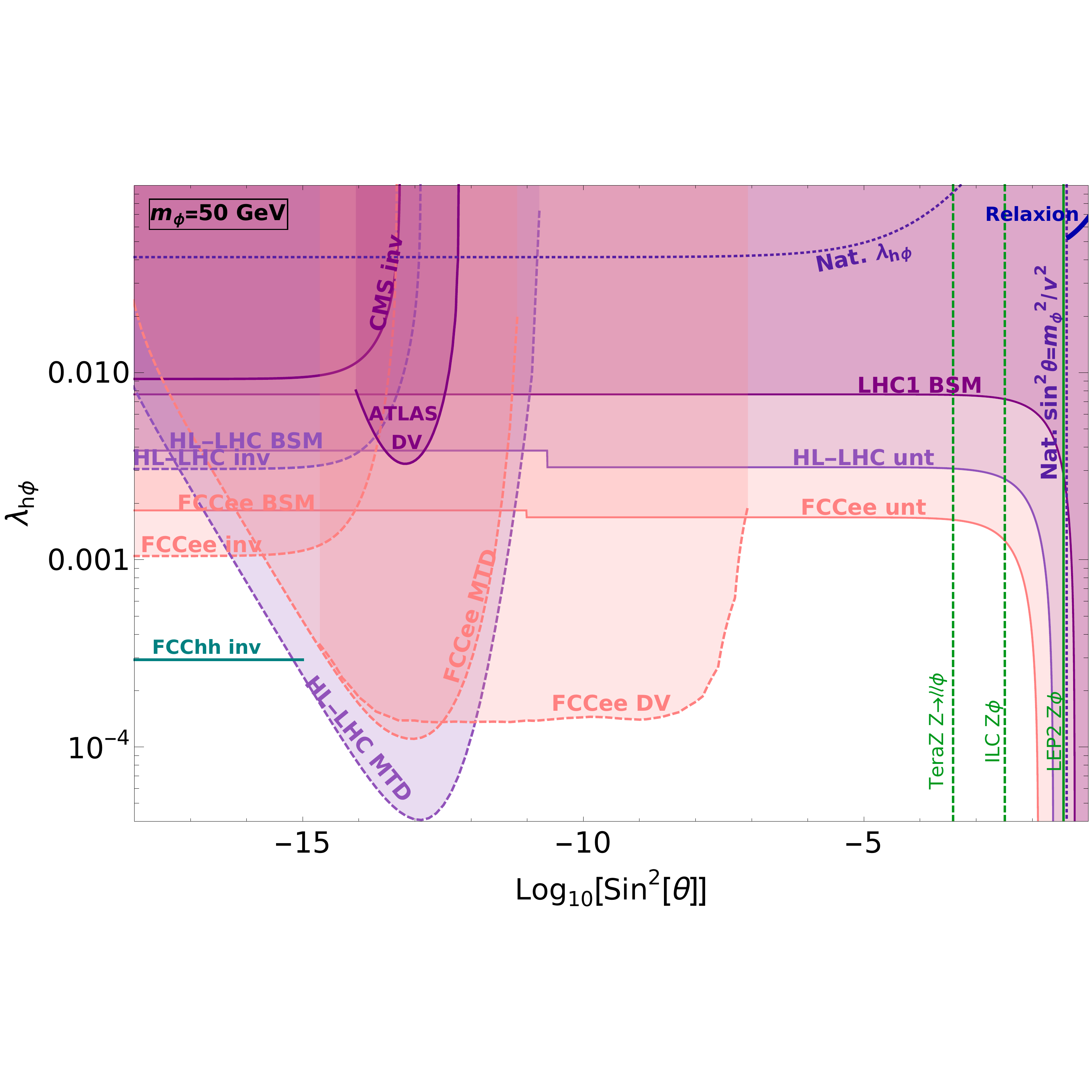}
		\caption{Direct and indirect bounds on $\lambda_{h\phi}$ and $\st$ for the generic
                  scalar singlet with a mass of $m_\phi=\{5,\,25,\,50\}\gev$, respectively.  The
                  dotted lines represent the upper naturalness bounds on $\lambda_{h\phi}$ and
                  $\sin\theta$ for the singlet.  The thick blue lines represent the viable relaxion
                  parameter space.  In the $m_\phi=50\gev$ plot, we use the result of the ATLAS
                  search for displaced jets~\cite{Aaboud:2019opc} for $m_\phi=55\gev$.  }
		\label{fig:portalAll}
	\end{center}
\end{figure}

In Fig.~\ref{fig:portalAll} we show the
coupling parameter plane spanned by $\sin^2\theta$ and \lhp{} for benchmark values of $m_\phi = \{5,\,25,\,50\}\gev$. For the \emph{singlet}, the upper
bound on \lhp{} from naturalness according to Eq.~\eqref{eq:lhphimax} is shown as the dotted curve.
The dotted vertical line represents the natural upper bound on $\sin^2\theta$, see
Eq.~\eqref{eq:genmixmax}. In contrast, for the \emph{relaxion}, the accessible \lhp{} within the
natural band of $\sin^2\theta$ is confined to the dark blue line that extends to larger
$\sin^2\theta$ than in the renormalizable singlet model, see Fig.~\ref{fig:RelaxionSpace_sth_lhp_m}.

For both models, \lhp{} only impacts the decay of the Higgs into a pair of singlets, \ie~the number
of produced $\phi$s, whereas $\sin\theta$ mainly determines their lifetime $\tau_\phi$, and only
contributes to $\br_{h\to\phi\phi}$ for high $\sin\theta$.

The bounds from direct searches for invisible Higgs decays form horizontal, almost mass-independent,
asymptotes on \lhp{} for sufficiently small $\sin^2\theta$, where a scalar of the considered mass is
still long-lived. Around this mass-dependent endpoint, the limit quickly weakens into a vertical
asymptote. Both the analyses of displaced vertices and the timing method probe several orders of
magnitude of $\sin^2\theta$. The reach in $\lambda_{h\phi}$ of the ATLAS \ac{DV} search is the strongest
for an intermediate mass of $m_\phi=25\gev$, and relatively mass-independent at the FCCee, whereas
the timing bounds become stronger for higher masses.  Here we show the bounds on the untagged Higgs
decays introduced in Sec.~\ref{sec:untagged} only for large enough values of $\sin^2\theta$, to
ensure a decay within the detector. For smaller $\sin^2\theta$, we show instead the (weaker) bounds
on the additional Higgs width $\Gamma_\bsm = \Gamma(h\to\phi\phi)$, that are valid regardless of the
decay length of $\phi$, hence down to arbitrarily low values of $\sin^2\theta$. Because the specific
decay of $\phi$ does not play a role, the shape is entirely determined by the \lhp{} and
$\sin\theta$ contributions to the coupling $c_{h\phi\phi}$ in $\br_{h\to\phi\phi}$. The green
vertical lines represent the LEP1 bound~\cite{Acciarri:1996um} for the rare $Z\to \phi \ell\ell$
decay, and the GigaZ and TeraZ projections we obtained by rescaling with the ratio of produced $Z$
bosons, or the bound on $e^+e^-\to Z\phi$ at LEP2~\cite{Schael:2006cr} and
ILC~\cite{Strategy:2019vxc} which are stronger than the respective $Z$-decay constraint for
$m_\phi = 50\gev$~\cite{Frugiuele:2018coc}.

The natural parameter space of the general singlet with $m_\phi=5\gev$ has not been probed yet. Only
small fractions of it can be probed by timing and displaced searches, as well as by fitting the untagged
and BSM Higgs width and by searches of rare $Z$-decays. For the higher masses considered, all
investigated bounds contribute to probing the natural parameter space, mainly because the upper
naturalness bounds increase with the mass.

Considering the relaxion at $m_\phi=5\gev$, so far only the $Z$-decays at LEP1 marginally constrain
the upper end of the natural region, which can be further probed by the same process at GigaZ, and
excluded by TeraZ. Furthermore, untagged Higgs decays at future colliders are sensitive to the natural
relaxion parameters. The heavier relaxion examples are already excluded by the \ac{BSM} Higgs decays at the 
LHC1.

\begin{figure}[tb]
  \begin{center}
    \includegraphics[trim={0 100pt 0 90pt}, clip,
    width=0.75\textwidth]{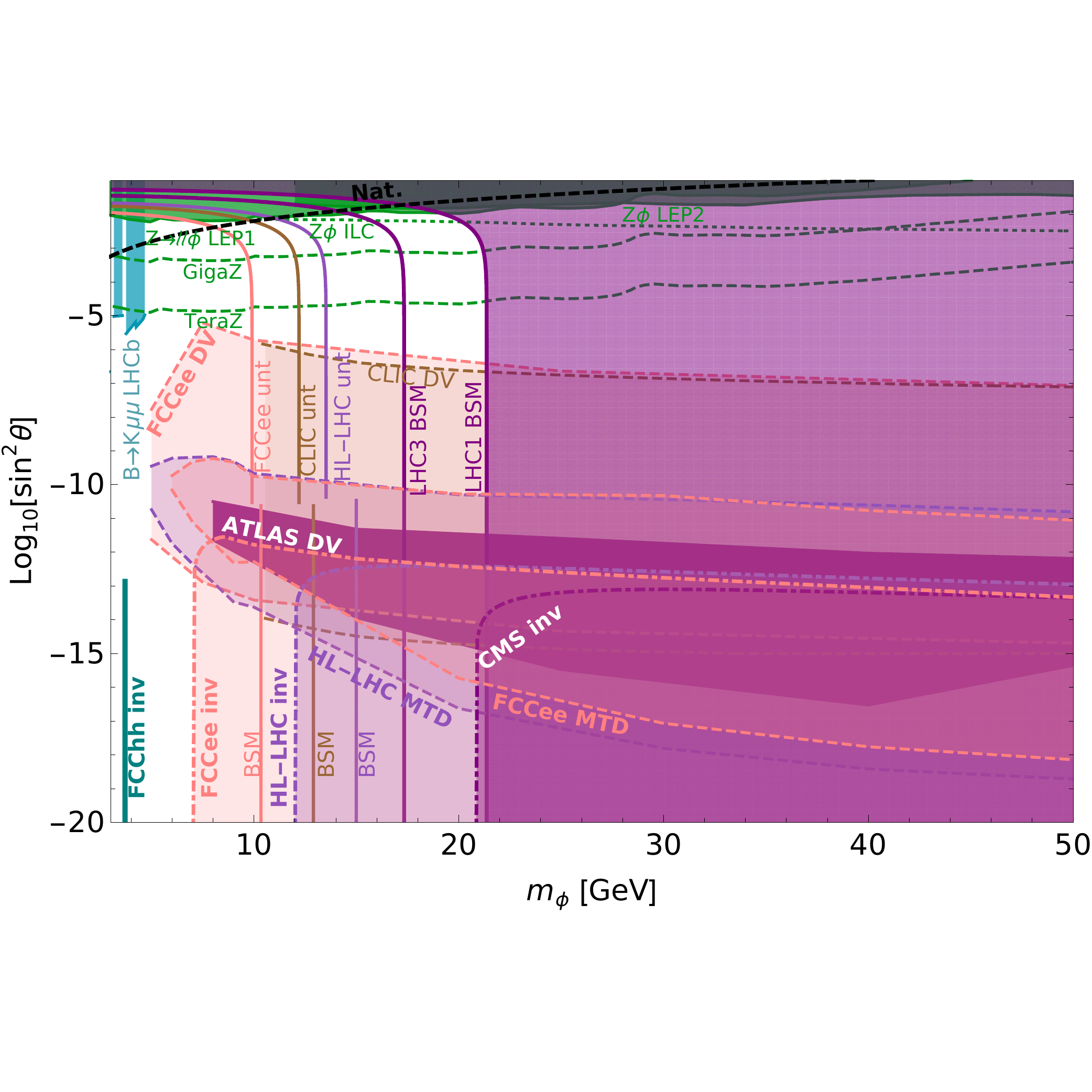}
    \caption{Bounds on $\st$ and $\mphi$ for the scalar singlet, with $\lhp{}=m_\phi^2/v^2$ stemming
      from various hadron and lepton colliders and covering a large range of life times. The bounds
      labeled by BSM arise from the collider indicated by the untagged bound of the same color. For
      the bounds in the prompt region see also Refs.~\cite{Flacke:2016szy, Frugiuele:2018coc}.}
    \label{fig:overview}
  \end{center}
\end{figure}

\begin{figure}[tb]
	\begin{center}
		\includegraphics[trim={0 89pt 0 90pt}, clip, width=0.75\textwidth]{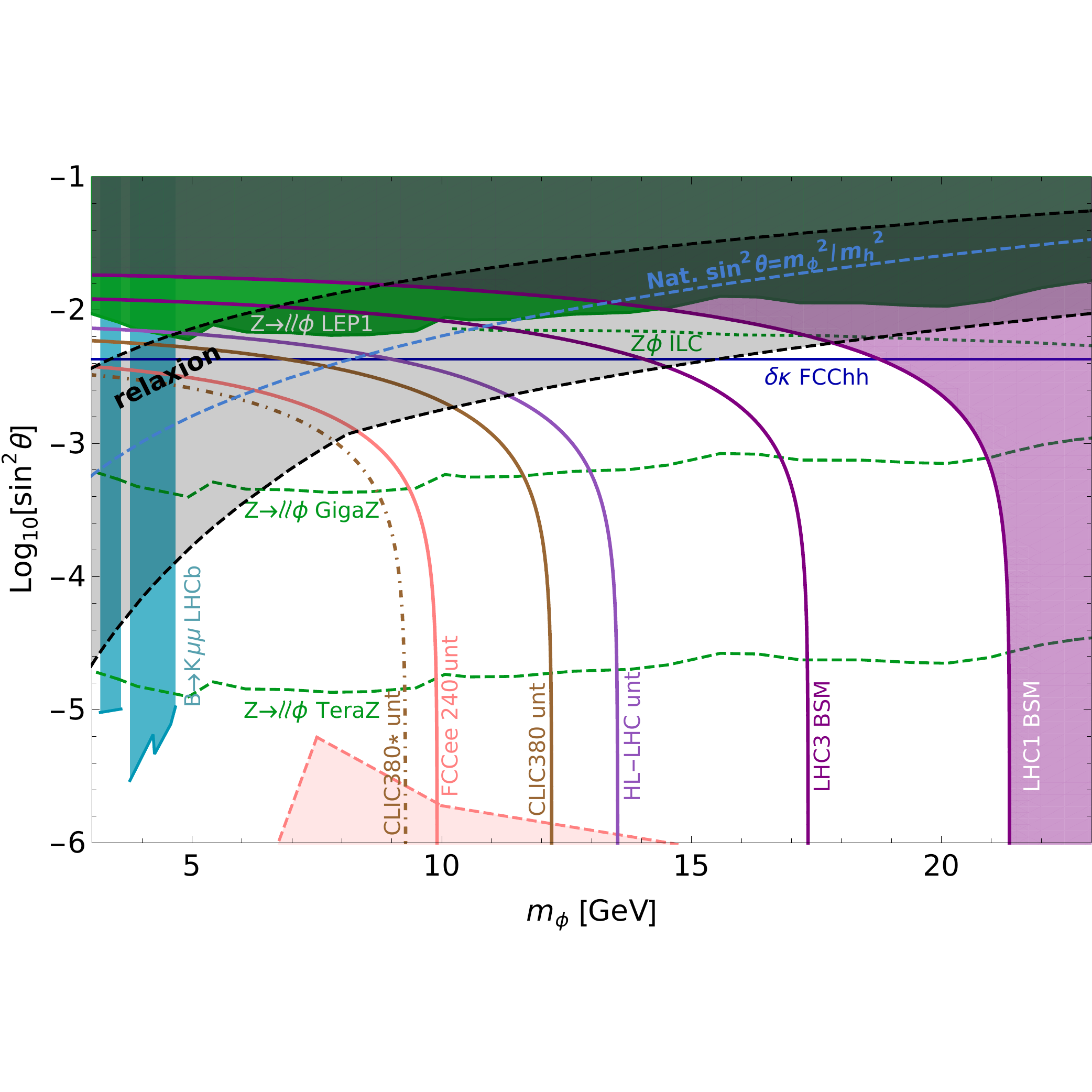}
		\caption{Prompt bounds on $\st$ and $\mphi$ for the relaxion, arising from direct
                  and indirect probes at various hadron and lepton colliders, as in
                  Fig.~\ref{fig:overview}. The dash-dotted line stems from the one-parameter fit for
                  CLIC as detailed in Sec.~\ref{sec:untagged}. The gray band marks the region where
                  the relaxion can be natural, see Sec.~\ref{sec:relaxion-as-special}. The upper
                  bound is given by Eq.~\eqref{eq:mixmax} for the relaxion stopping in the first
                  minimum ($n=1$). The lower bound for low $m_\phi$ is dominated by the stopping in
                  a generic minimum ($n \gg 1$), see Eq.~\eqref{eq:genmixmin}; for high $m_\phi$ by
                  the solution for $n=1$, see Eq.~\eqref{eq:mixmin}. The bands corresponds to a
                  choice of $\Lambda_\text{min}=\Mt_\textnormal{max} = 1\tev$.}
		\label{fig:combinedboundsRelevant}
	\end{center}
\end{figure}

In Figs.~\ref{fig:overview} and \ref{fig:combinedboundsRelevant} we show the bounds in the
$m_\phi$-$\sin^2\theta$ plane for the singlet scalar and for the relaxion, respectively. For the singlet
scalar, we set the coupling $\lhp =m_\phi^2/v^2=\hat{\lambda}_{h\phi}^\textnormal{max}$, hence \lhp{}
could be even larger. For the relaxion, the value of $\lhp$ is given by Eq.~\eqref{eq:lhpRel}. In
addition to the bounds discussed above, we also show the direct bound for $m_\phi < 5\gev$ from
$B\to K\mu\mu$ at the LHCb \cite{Aaij:2012vr,Aaij:2015tna,Flacke:2016szy}.  Furthermore, we
translate the uncertainties $\delta\kappa$ of the Higgs coupling modifier in global
fits\footnote{We obtain the approximate 95\% CL bound on $\sin^2\theta$ from the provided 68\% CL
  bound on $\delta\kappa$ with $\kappa = 1 +\delta\kappa$ by
  \mbox{$\sin^2\theta^{(95)} \simeq 1 - (1+r\,\delta\kappa^{(68)})^2$} where
  $r= \sqrt{q^{(95)}_n/q^{(68)}_n}$, and $q_n$ are the respective quantiles of a
  $\chi^2$-distribution with $n$ parameters.  } into model-independent bounds on $\sin^2\theta$
that are independent of $m_\phi$ and $\lambda_{h\phi}$. The strongest bound stems from
$\delta\kappa_Z$ at the FCChh (see Tab.~\ref{tab:HGfit}), and is shown in
Fig.~\ref{fig:combinedboundsRelevant}, but omitted in Figs.~\ref{fig:portalAll} and
\ref{fig:overview}.  From Fig.~\ref{fig:combinedboundsRelevant} we see that relaxions heavier than
$\sim18\gev$ are already excluded by the current LHC bounds on BSM Higgs decays. Rare $Z$-decays from
LEP1 probe parts of the natural parameter space of the relaxion for $m_\phi\gtrsim 5\gev$, but the
bound from the BSM Higgs branching at the LHC Run-1 is stronger than this LEP1 bound for
$m_\phi\gtrsim 15\gev$. The best bounds from untagged Higgs decays will come from the FCChh, and can
exclude relaxions above $m_\phi\gtrsim 8\gev$. On top of that, TeraZ can exlude relaxions of
$m_\phi\gtrsim 3\gev$.

\section{Conclusions}
\label{sec:conclusion}
In this work, we exploit the sensitivity of the exotic Higgs decay channel $h\to\phi\phi$ to
parameters of the relaxion and singlet models, taking into account existing searches and global fits,
as well as projections for future colliders.

We discuss the renormalizable, non-$\mathds{Z}_2$-symmetric singlet extension of the \ac{SM},
focusing on the exotic Higgs decay $h\to\phi\phi$ via the triple scalar coupling
$c_{h\phi\phi}$. The collider phenomenology is determined by the four parameters $m_\phi$,
$\sin\theta$, $\hat\lambda_{h\phi}$, and $a_{\phi}$. Beyond the usual naturalness bound on the
mixing angle, we present naturalness bounds on $\hat{\lambda}_{h\phi}$ and $a_\phi$ and investigate
their implication on the physical parameter space.  Moreover, we provide a matching between the
singlet parameters and those of the relaxion. Here, the absence of a $\mathds{Z}_2$ symmetry is pivotal to
accommodate the linear slow-roll relaxion potential. The $h^2\phi^2$ term in the singlet model maps
onto the first term of the expansion of the backreaction potential.  We extend the naturalness
relaxion band to higher masses relevant at colliders, where it is described by only two parameters,
$m_\phi$ and $\sin\theta$, which determine \lhp{}. Consequently, the relaxion model is both more
constrained and predictive than the renormalizable singlet extension. 

The lifetime of $\phi$, given by $\sin\theta$ and $m_\phi$, is the crucial handle in determining the kind of search strategy that sets the strongest bound. We study various lifetime dependent strategies. In particular, we evaluate the limits from global coupling fits on the new Higgs branching ratio into \ac{BSM}, split into untagged and invisible final states; interpret the searches for the Higgs decaying into displaced jets in terms of the singlet model; exploit the time delay of jets originating from the $\phi$ decay to derive bounds in the region of intermediate lifetime; constrain the region of low $\sin^2\theta$ by searches for invisible Higgs decays and pay attention to the range where the decay lengths are of the detector size such that only a fraction of the particles actually gives rise to the invisible signature.

Our main phenomenological findings are:
\begin{itemize}
\item For $m_\phi=5\gev$, only a small fraction of the natural singlet parameter space can be
  probed. For higher masses, larger coupling values become natural and the LHC has already excluded
  parts of it.
\item The FCC can probe almost the complete considered parameter region by combining TeraZ, FCCee
  and FCChh, unless \lhp{} is much smaller than used here.
\item The natural range for relaxions heavier than $18\gev$ is already excluded by searches for
  untagged Higgs decays at the LHC. The FCCee has the potential to exclude relaxions down to $8\gev$
  using the same strategy. Only the search for rare $Z$ decays at TeraZ will be able to exclude the
  full mass range for heavy relaxions with $m_\phi>3\gev$.
\end{itemize}

\section*{Acknowledgments}
We thank Gilad Perez for the initial collaboration, many valuable discussions, and helpful comments
on the draft.  Moreover, we thank Zhen Liu for useful information on timing bounds, and Christophe
Grojean for interesting discussions. E.F.~acknowledges the support by the Minerva Foundation during
the initial phase of this work. M.S.~is supported by the Alexander von Humboldt Foundation and
thanks Fermilab for hospitality. The work of O.M.~is supported by the Foreign Postdoctoral
Fellowship Program of the Israel Academy of Sciences and Humanities. I.S.~is supported by a fellowship from the Ariane de Rothschild Women Doctoral Program.

\markboth{}{}

\appendix

\section{Relaxion stopping point}
\label{sec:relaxstop}
The backreaction and the slow-roll potentials are defined in Eqs.~\eqref{eq:br} and \eqref{eq:Vsr},
respectively. Here we only consider $j=2$.  The relaxion stops its evolution at
$\phi_0 \equiv \theta_0 f$, given by
\begin{equation}\label{eq:stab}
  V'_{\text{br}}(\theta_0) = - V'_{\text{sr}} \quad \Longrightarrow \quad  \frac{v^2(\phi_0)  \tilde M^2}{2 f} \sin \theta_0 = - g \Lambda^3\,,
\end{equation}
In the following we set $r=1$ for simplicity, given that the exact expression for the relaxion mass
only mildly depends on $r$.  As the relaxion rolls down its potential before stopping, during each
relaxion period $\Delta \phi =2 \pi f$ the maximal (in absolute value) slope of the oscillatory
potential $V'_{\text{br}}$ changes by
\begin{equation}
  \Delta V'_{\text{br}} 
  \;\simeq\; \frac 1 {2f} \Delta v^2 \tilde M^2 \sin \theta_\star 
  \;\simeq\;  \frac{\pi}{\lambda_h} g \Lambda \tilde M^2 \sin \theta_\star 
  \;\simeq\; - \frac{\pi}{\lambda_h r} \frac{\tilde M^2}{\Lambda^2} \frac{v^2(\phi_0) \tilde M^2}{2 f} \sin \theta_0 \sin \theta_\star\,,
\end{equation}
where $\theta_\star = \phi_\star/f$ denotes the relaxion angle at which the $V_{\text{br}}$ slope is
maximized within the given $2 \pi f$ period, \ie{}~the inflection point of the periodic potential.
$\Delta V'_{\text{br}}$ is $\tilde M^2/\Lambda^2$ suppressed with respect to the $V'_{\text{br}}$
overall size at the stopping point~(\ref{eq:stab}).  Close to the final minimum, $\theta_\star$ can
be found using Eq.~\eqref{eq:relaxion_pheno_from_theory_parameters_mphi} for
$m^2_\phi \simeq V''(\phi_\star)$, and solving
\begin{equation}
V''(\phi_\star) = 0 \quad \Longrightarrow \quad \frac {\cos \theta_\star}{\sin^2 \theta_\star} 
= \frac {\tilde M^2}{\lambda_h v^2(\phi_\star)}\,,
\end{equation}
where
$\lambda_h v^2(\phi_\star) \simeq \lambda_h v^2(\phi_0)+\tilde M^2 \sin \theta_\star
(\theta_0-\theta_\star)$ by a Taylor expansion of $\mu^2(\phi)$ neglecting the term suppressed by
$g$.  After the first minimum is formed, the slope of the periodic potential can overcompensate
$V'_{\text{sr}}$ only by $\Delta V'_{\text{br}}$. After the $n$-th minimum it can do so by
$n \, \Delta V'_{\text{br}}$. Correspondingly, the slope of the overall potential is given by the
same value
\begin{equation}
V'(\phi_\star) \simeq n \, \Delta V'_{\text{br}}\,.
\end{equation} 
We therefore know the position of the inflection point $\phi_\star$ and its slope
$V'(\phi_\star)$. They can be used to find the properties of the closest minimum $\phi_0$ located
before $\phi_\star$. The value of $V^\prime (\phi_0)$ can be expressed as a Taylor series around
$\phi_\star$
\begin{equation}\label{eq:vprimeexp}
0=V'(\phi_0) = V' (\phi_\star) + \frac{1}{2}V'''(\phi_\star) (\phi_0 - \phi_\star)^2 + \dots 
\end{equation}
with
\begin{equation}
  V''' \simeq - \frac{\tilde M^2}{2 f^3} \left( \frac 3 {2\lambda_h} \tilde M^2 \sin 2\theta_\star + v^2(\phi_\star) \sin \theta_\star  \right)\,.
\end{equation}
Note that $V'''$ is obtained from the effective relaxion potential $V_\textrm{eff}$ after
integrating out the Higgs boson, $h^2 \to -\mu^2(\phi)/\lambda_h$, which is given by
\begin{equation}
V_\textrm{eff}= -\frac{1}{4 \lambda_h} \mu^4(\phi) + V(\phi)\,,
\end{equation}
with $\mu^2(\phi_0)=-\lambda_h v^2$.
Eq.~(\ref{eq:vprimeexp})  allows to find $\phi_0$ from
\begin{equation}\label{eq:deltaphi}
(\phi_0 - \phi_\star)^2 \simeq  - 2 V' (\phi_\star) / V'''(\phi_\star)\,,
\end{equation}
and consequently all the related parameters of the theory. In particular, the relaxion mass can be
approximated as
\begin{equation}
m_\phi^2 = V''(\phi_0) \simeq V'''(\phi_\star) (\phi_0 - \phi_\star) \simeq \sqrt{|2 V' (\phi_\star) V'''(\phi_\star)|}\,.
\end{equation}
As we see, the mass is proportional to $\sqrt{V' (\phi_\star)}$, which itself carries a factor
$\tilde M/\Lambda$. This is precisely the reason why the relaxion mass is suppressed with respect to
the naive estimate.

In this paper, we are interested in the corner of the parameter space where the relaxion reaches its
maximal possible masses, which requires taking $\tilde M \gtrsim v$. In the limit
$\tilde M \gg \sqrt{\lambda_h}v$, applicable within the relaxion mass range considered in this work,
the relevant expressions simplify to
\begin{eqnarray}
\theta_0 &\simeq& - \frac {\sqrt{\lambda_h} v} {\tilde M} + \sqrt n \sqrt{\frac {3\pi \lambda_h^{1/2}} {2}} \frac{v^{3/2}}{\Lambda \tilde M^{1/2}}\,, \\
m_\phi^2 & \simeq& \sqrt n \sqrt{\frac{3 \pi}{2 \lambda_h^{1/2}}} \frac{(v \tilde M)^{5/2}}{f^2 \Lambda} \label{eq:highmass}\,.
\end{eqnarray}
Inserting the relaxion angle $\theta_0$ into the general expression for the relaxion mass in
Eq.~(\ref{eq:relaxion_pheno_from_theory_parameters_mphi}), we see that the small relaxion mass
appears as a result of a fine cancellation between two contributions.  Note that this also means
that the loop corrections, otherwise subleading, may contribute sizeably to the relaxion mass. This,
however, should not change qualitatively the results that we have derived, as the presence of the
relaxion mass suppression is directly linked to the slow growth of the periodic barriers
amplitude---$\Delta V'_{\text{br}}/V'_{\text{br}}\ll1$---the feature which is not expected to be
altered by the loop effects.

For completeness we also write down corresponding expressions in the opposite limit,
$\tilde M \ll v$, relevant for lighter relaxion, which were derived in
Ref.~\cite{Banerjee:2020kww}\footnote{The reversed sign of $\theta_0$ is a consequence of a
  different sign convention for the relaxion potential.}
\begin{eqnarray}
  \theta_0 &\simeq& -\pi/2 + \frac {\tilde M^2}{\lambda_h v^2} + \sqrt n \sqrt{\frac{2 \pi}{\lambda_h}}\frac{\tilde M}{\Lambda}\,, \\
  m_\phi^2 & \simeq & \sqrt n \sqrt{\frac{\pi}{2 \lambda_h}} \frac{v^2 \tilde M^3}{ f^2 \Lambda}\,.
\end{eqnarray}

\section{Estimating singlet production via Higgs mixing}
\label{sec:production-via-higgs}

For small values of the coupling $\lambda_{h\phi} \ll \sint^2 m_h^2/(2v^2)$, the branching ratio
$\br_{h\to\phi\phi}$ is proportional to $\sin^4\theta$,
cf.~Eq.~\eqref{eq:chphiphi_singlet_approx2}. If in addition $\sin\theta$ is small, the Higgs almost
never decays into a pair of scalars. On the other hand, the production of scalars via their mixing
with the Higgs only scales as $\sin^2 \theta$ and becomes the dominant production mechanism if
\lhp{} is small. However, if a sufficiently long lifetime is required in order to have a handle for
the considered analyses, we estimate in the following that production via mixing yields only few
events making a dedicated search difficult.

The number of scalars produced via mixing is given by
$n_\textnormal{mix} = \mathcal{L} \sigma_\phi \sint^2$, where $\mathcal{L}$ is the luminosity and
$\sigma_\phi$ is the production cross section of a Higgs boson with mass equal to $m_\phi$. Since
detecting a dijet resonance at low mass is extremely challenging, we will consider only the searches
for displaced jets or missing energy. To obtain a displaced or invisible signature, we need
$c\tau \gtrsim 1\,\textnormal{cm}$ $(\gtrsim 1\,\mu\textnormal{m})$ for the HL-LHC (FCCee) which
translates into \mbox{$\sin^2\theta \lesssim 10^{-9}$} \mbox{$(\lesssim 10^{-5})$} using
Fig.~\ref{fig:ctau} for $m_\phi=5\gev$. A higher value for $m_\phi$ would be helpful in an analysis,
but at the same time require even smaller mixing angles and also imply a smaller production cross
section for kinematical reasons.

The HL-LHC will collect a luminosity of $\mathcal{L} = 3 \cdot 10^{6}\pb^{-1}$. The production cross
sections for a light Higgs at the LHC are below $100\pb$ for all modes except for gluon fusion
without $p_T$ requirement~\cite{Frugiuele:2018coc}. A leading order parton-level estimate with
MadGraph5 for $\phi + j$ production at $14\tev$ with a very mild $p_T>20\gev$ requirement for the
scalar yields $\sigma_\phi\approx 120\pb$. Therefore the HL-LHC can only produce
$n_\textnormal{mix}^\textnormal{HL-LHC} \lesssim 0.4$ scalars.  Consequently, even before
significant selection cuts no events will be available for an analysis.

The FCCee on the other hand will collect $\mathcal{L}=5\cdot 10^6\pb^{-1}$ and the dominant
production mode for a light Higgs at FCCee is Higgs-strahlung with a cross section of about $0.6\pb$
\cite{Frugiuele:2018coc} with $p^\phi_T> 10\gev$.  Therefore
$n_\textnormal{mix}^\textnormal{FCCee}\lesssim 30$.  Considering more selective cuts on top of the
minimal example cut applied here as well as the detector acceptance and \eg{} leptonic $Z$ decay
branching ratios, it will be impossible to have a sufficient number of scalars left for an analysis.

Here we argued why we focus only on $\phi$ production via Higgs decays. However, progress in
detecting promptly decaying low-mass resonances may provide a new channel for singlet and relaxion
searches \cite{Frugiuele:2018coc}. Especially $bb$, $\tau\tau$ or $\mu\mu$ decays from production
via mixing may allow to constrain the parameter regions where \lhp{} is negligible.

\section{Timing of delayed jets}\label{appendix:timing}
The crucial requirement of the analysis proposed in Ref.~\cite{Liu:2018wte} is that a jet leaving no
track in the inner tracker hits the proposed timing layer with a delay $\Delta t>1\ns$ with respect
to a (hypothetical) \ac{SM} jet, going directly from the interaction point to the same location
on the timing layer. This signature can be achieved by a particle that is invisible to the inner
detector and decays into \ac{SM} hadrons between the inner tracker and the timing layer. The delay
then is a result both of the lower velocity of the heavier decaying particle, and of the displacement of the secondary decay in which the hadron is produced. For this reason, the acceptance probability of a given event is dominated by the geometrical trajectory of the decaying scalar and its decay product, once the kinematics is determined. Namely, once the four-momenta of the scalar and the jet are set, the lab-frame decay length of the scalar determines the secondary vertex position, the position in which the final jet hits the timing layer, and the overall time delay.

Since the analysis only requires at least one delayed jet, we can consider the four final state jets
from the decay chain $h\to\phi\phi\to4 j$ independently. Then, for a jet in a given event, we can
find the range of allowed lab frame decay lengths of the scalar $l_{\phi, j}$, for which the jet will be accepted as signal. If this range is non-empty, we can assign a weight $w_j$, calculated as the
probability to obtain $l_{\phi, j}$ within the allowed range, given that the proper decay length is
$c\tau_\phi$, as in Eq.~\eqref{eq:w}.
The probability for the whole event to be accepted is then given by
$\epsilon_\text{event} = 1-(1-w_1)(1-w_2)(1-w_3)(1-w_4)$.

In the following we will explain the computation of the allowed range of $l_{\phi, j}$. As described
above, the scalar needs to decay between the outer radius of the inner tracker $L_1$ and the outer
radius of the timing layer $L_2$. For CMS $L_1=0.2\m$ and $L_2=1.17\m$~\cite{Liu:2018wte}, and for the FCCee we assume
$L_1=0.127\m$ and $L_2=2.1\m$~\cite{Bacchetta:2019fmz}. Thus, the distance the scalar may travel before decaying is constrained by $ l^{L_1}\leq l_{\phi,j}\leq l^{L_2}$, given by
\begin{align}
l^{L_1}=\frac{L_1}{\sin\theta_\phi}\,,\qquad 
l^{L_2}=\frac{L_2}{\sin\theta_\phi}\,,
\end{align}
where $\theta_\phi$ is the polar angle between the beam axis and the three-momentum of the
considered scalar. In addition, we demand that the displaced jet does not cross the inner radius $L_1$ towards the beam axis, as it will leave a signature in the tracker. We thus require
\begin{align}
l_{\phi, j}^\textnormal{min} &= \frac{L_1}{\sin\theta_\phi}\textnormal{max}\left( 1, - \frac{\textnormal{sign}(\cos\varphi)}{|\sin\varphi|}\right)\,,
\end{align}
where $\varphi\equiv \varphi_\phi-\varphi_j$, and $\varphi_\phi$ and $\varphi_j$ refer to the azimuthal angles of the scalar and the jet, respectively.

The main selection criterion of the search is the time delay of the decay product, which is a result of the displaced vertex. The delay is defined as
\begin{align}
\Delta t&= \frac{l_\phi}{c\beta_\phi}+\frac{l_j}{c\beta_j}-\frac{l_\text{SM}}{c\beta_\text{SM}}~\label{eq:tdelay}\,,
\end{align}
where $l_\phi$ is the distance traveled by the scalar before it decays, $l_j$ is the distance
traveled by the decay product (a jet, in our case) to the timing layer, and $l_\text{SM}$ is the
distance a hypothetical \ac{SM} particle would travel directly from the interaction point to the timing
layer. The velocities of the particles are denoted by $\beta_\phi$, $\beta_j$ and $\beta_\text{SM}$
in units of the speed of light $c$. Because the \ac{SM} hadrons are light, $\beta_\text{SM}=1$ to a good
approximation. By demanding that the delayed jet hits the timing layer at radius $L_2$, and by setting $l_\text{SM}=|\vec{l}_\phi+\vec{l}_j|$, the time delay can be expressed solely as a function of the event kinematics and $l_\phi$. As the time delay has at most one maximum as a function of $l_\phi$, the allowed decay should lie between $l^{\Delta t}_\text{max}, l^{\Delta t}_\text{max}$, given by solving Eq.~\eqref{eq:tdelay} for $l_\phi$ with the required minimal time delay. Note that Eq.~\eqref{eq:tdelay} can be brought to a 4th-degree polynomial form in $l_\phi$, and thus its roots can be found analytically. The temporal resolution of the timing layer is simulated by assigning normally distributed time stamps to the displaced jet hit $\delta_j$ and to the \ac{SM}-\ac{ISR} hit $\delta t_\text{ISR}$, smeared by $\sigma=30$ ps~\cite{timingDetectors}, and requiring $\Delta t_\text{th}\leq\Delta t+\delta t_j-\delta t_\text{ISR}$, where $\Delta t_\text{th}=1$~ns is the minimal time delay set by the analysis.

Lastly, the decay product of the scalar should hit the timing layer at $L_2$ within the length of the detector (in the $\hat{z}$ direction), where we set $|z_\text{max}|=2.6\m$ at CMS and $|z_\text{max}|=2.3\m$ at the FCCee. If the scalar is produced at $z_0$, then the $z$ position of the hit of its decay product is 
\begin{align}
Z&\equiv l_\phi\cos\theta_\phi+l_j\cos\theta_j-z_0\,,
\end{align}
which is yet again completely determined by the event kinematics and $l_\phi$ (we set $z_0=0$ for simplicity, as the variations in the exact primary vertex position are negligible compared to the detector length). Therefore, imposing $-|z_\text{max}| \leq Z\leq |z_\text{max}|$ and solving for $l_\phi$ yields another set of constraints. Note that since $Z$ can have at most one extremum as a function of $l_\phi$, there may be at most two disconnected allowed ranges of $l_\phi$ satisfying the requirement above.

\begin{figure}[ht!]
	\begin{center}
		\subfloat[ \ac{HL}-\ac{LHC} $\sqrt{s}=14$~TeV.]{ \includegraphics[scale=0.5]{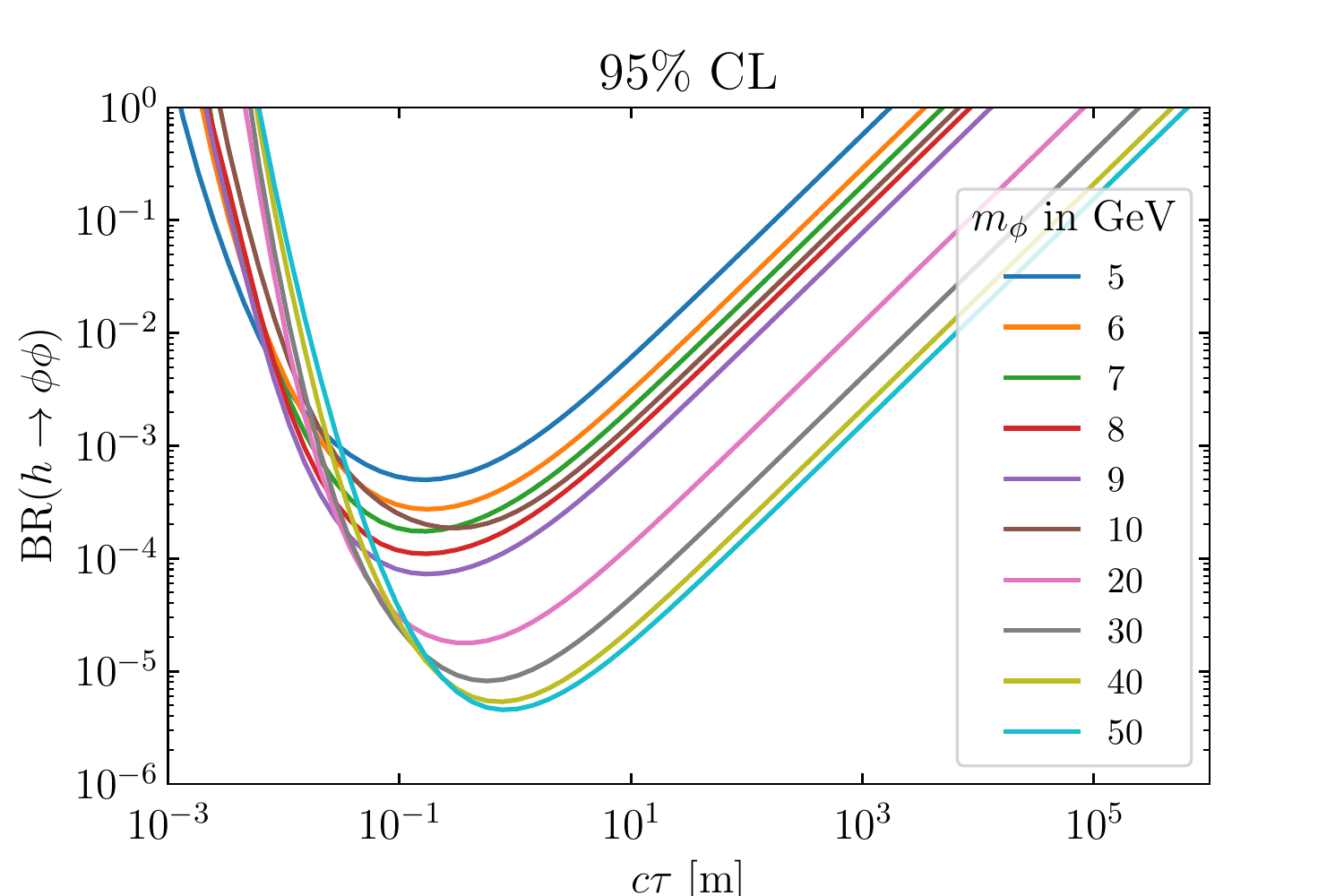}}	
		\subfloat[ FCC-ee $\sqrt{s}=240$~GeV.]{\includegraphics[scale=0.5]{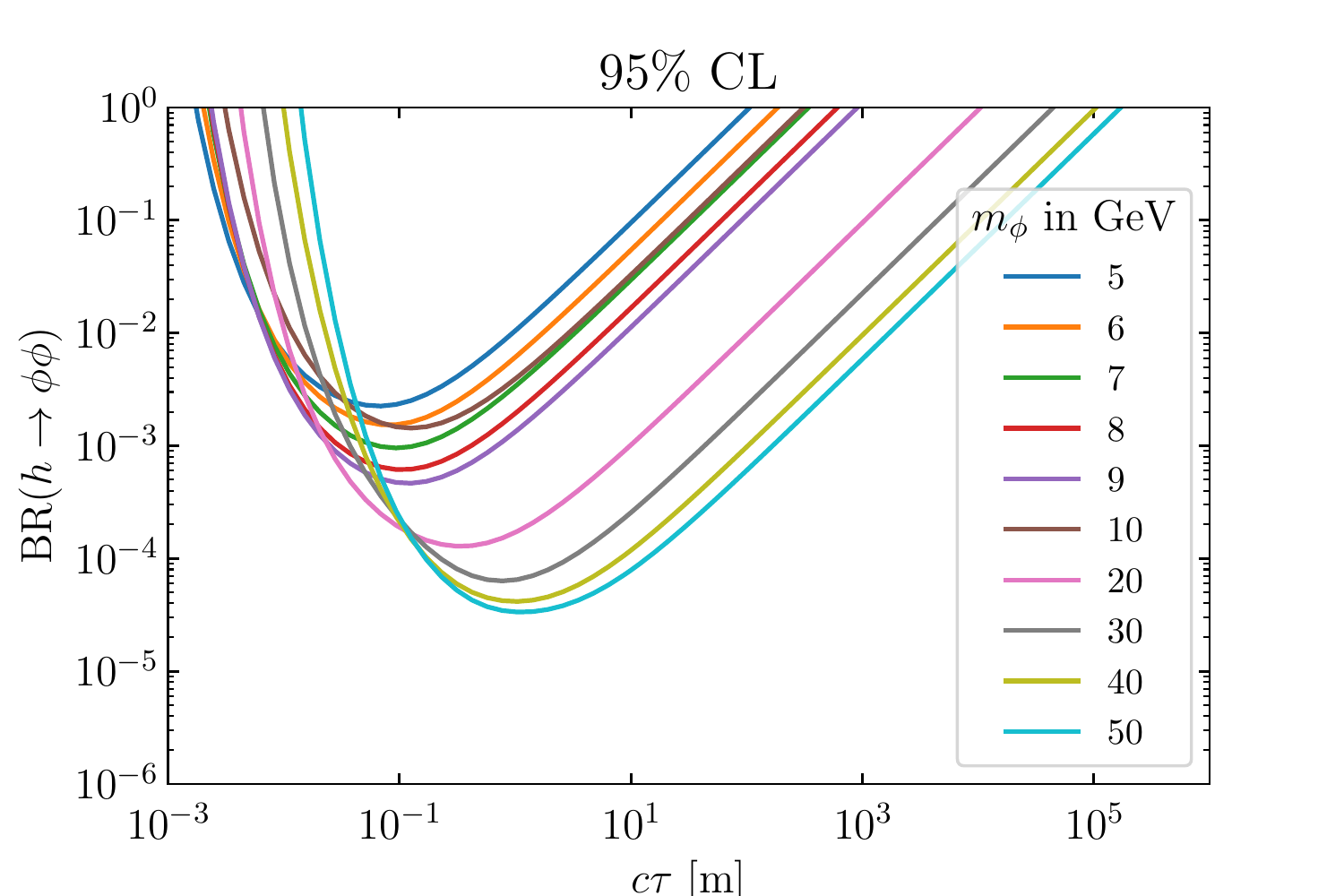}}\\
		\subfloat[\ac{HL}-\ac{LHC} $\sqrt{s}=14$~TeV.]{ \includegraphics[scale=0.5]{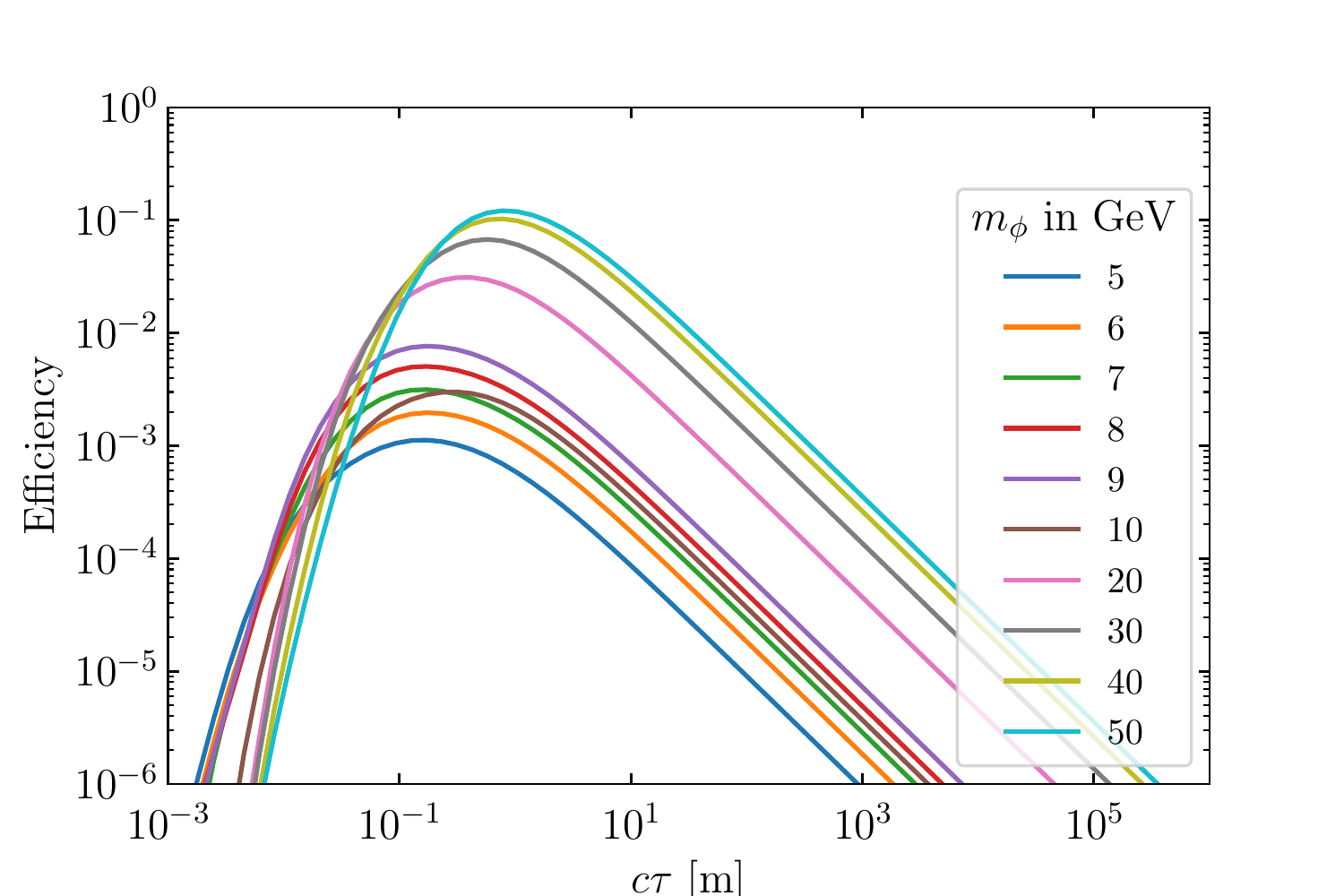}}
		\subfloat[FCC-ee $\sqrt{s}=240$~GeV.]{\includegraphics[scale=0.5]{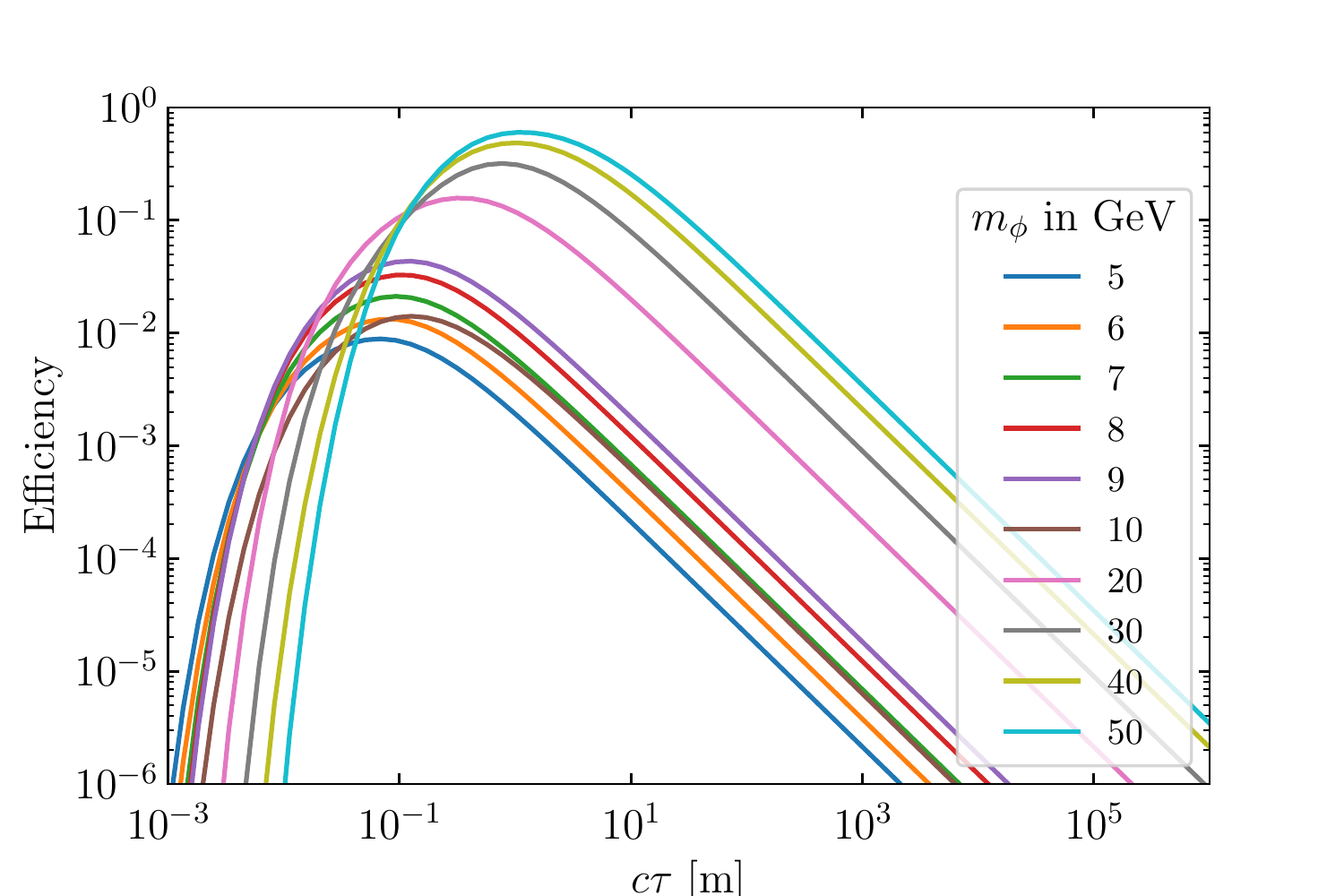}}
		\caption{$\br(h\to\phi\phi)$ and efficiency as a function of $c\tau_\phi$ for a search for delayed jets.}
		\label{fig:Eff_vs_ctau}
	\end{center}
\end{figure}

The final range of allowed decay lengths is then set by the union of the constraints given by the conditions above. For each allowed continuous range of $l_\phi$, $w$ is calculated by
\begin{equation}
w_j=\exp\leri{-\frac{l_{\phi,j}^\text{min}}{c\tau_\phi\gamma_\phi\beta_\phi}}-\exp\leri{-\frac{l_{\phi,j}^\text{max}}{c\tau_\phi\gamma_\phi\beta_\phi}}\,.
\end{equation}
If the union has two or more disconnected regions, their contribution to $w$ should be summed. The resulting bounds on the Higgs branching to a pair of scalars and the efficiency of the search, both as a function of the lifetime, are presented in Fig~\ref{fig:Eff_vs_ctau}.

\label{app:efficiency}

\bibliographystyle{JHEP}
\bibliography{RelaxionCollider}

\end{document}
